\pdfoutput=1
\documentclass[12pt,a4paper]{article}

\usepackage{ifthen} 
\newboolean{pdflatex}
\setboolean{pdflatex}{true} 

\newboolean{articletitles}
\setboolean{articletitles}{true} 

\newboolean{uprightparticles}
\setboolean{uprightparticles}{false} 

\def\paperauthors{LHCb collaboration}
\def\paperasciititle{Observation of the B0 and Bs to Dstar0bar K pi decays} 
\def\papertitle{Observation of the $\Bz\to\Dstarzb\Kp\pim$ and $\Bs\to\Dstarzb\Km\pip$ decays}
\def\paperkeywords{{High Energy Physics}, {LHCb}}
\def\papercopyright{\the\year\ CERN for the benefit of the LHCb collaboration} 
\def\paperlicence{CC BY 4.0 licence}
\def\paperlicenceurl{https://creativecommons.org/licenses/by/4.0/}

\usepackage[top=1in, bottom=1.25in, left=1in, right=1in]{geometry}

\usepackage{microtype}
\usepackage{lineno}  
\usepackage{xspace} 
\usepackage{caption} 

\usepackage{graphicx}  
\usepackage{color}
\usepackage{colortbl}
\graphicspath{{./figs/}} 

\usepackage{amsmath} 
\usepackage{amssymb}
\usepackage{amsfonts}
\usepackage{upgreek} 

\newcommand*\patchAmsMathEnvironmentForLineno[1]{%
\expandafter\let\csname old#1\expandafter\endcsname\csname #1\endcsname
\expandafter\let\csname oldend#1\expandafter\endcsname\csname
end#1\endcsname
 \renewenvironment{#1}%
   {\linenomath\csname old#1\endcsname}%
   {\csname oldend#1\endcsname\endlinenomath}%
}
\newcommand*\patchBothAmsMathEnvironmentsForLineno[1]{%
  \patchAmsMathEnvironmentForLineno{#1}%
  \patchAmsMathEnvironmentForLineno{#1*}%
}
\AtBeginDocument{%
\patchBothAmsMathEnvironmentsForLineno{equation}%
\patchBothAmsMathEnvironmentsForLineno{align}%
\patchBothAmsMathEnvironmentsForLineno{flalign}%
\patchBothAmsMathEnvironmentsForLineno{alignat}%
\patchBothAmsMathEnvironmentsForLineno{gather}%
\patchBothAmsMathEnvironmentsForLineno{multline}%
\patchBothAmsMathEnvironmentsForLineno{eqnarray}%
}

\usepackage{hyperxmp}

\usepackage[pdftex,
            pdfauthor={\paperauthors},
            pdftitle={\paperasciititle},
            pdfkeywords={\paperkeywords},
            pdfcopyright={Copyright (C) \papercopyright},
            pdflicenseurl={\paperlicenceurl}]{hyperref}

\usepackage[colorinlistoftodos,textsize=scriptsize]{todonotes}

\usepackage[bottom,flushmargin,hang,multiple]{footmisc}

\usepackage[all]{hypcap} 

\usepackage{xspace}
\usepackage{upgreek}

\newcommand{\offsetoverline}[2][0.1em]{\kern #1\overline{\kern -#1 #2}}%

\def\lhcb   {\mbox{LHCb}\xspace}

\def\MagUp {\mbox{\em Mag\kern -0.05em Up}\xspace}

\ifthenelse{\boolean{uprightparticles}}%
{

 \def\Peta        {\ensuremath{\upeta}\xspace}

 \def\Ppi         {\ensuremath{\uppi}\xspace}

 \def\PDelta      {\ensuremath{\Delta}\xspace}
 \def\PXi         {\ensuremath{\Xi}\xspace}
 \def\PLambda     {\ensuremath{\Lambda}\xspace}
 \def\PSigma      {\ensuremath{\Sigma}\xspace}
 \def\POmega      {\ensuremath{\Omega}\xspace}
 \def\PUpsilon    {\ensuremath{\Upsilon}\xspace}

 \def\PB      {\ensuremath{\mathrm{B}}\xspace}
 
 \def\PD      {\ensuremath{\mathrm{D}}\xspace}

 \def\PK      {\ensuremath{\mathrm{K}}\xspace}

 \def\Pb      {\ensuremath{\mathrm{b}}\xspace}
 \def\Pc      {\ensuremath{\mathrm{c}}\xspace}

 \def\Pi      {\ensuremath{\mathrm{i}}\xspace}

 \def\Pp      {\ensuremath{\mathrm{p}}\xspace}

 \def\Ps      {\ensuremath{\mathrm{s}}\xspace}
 
 \def\Pu      {\ensuremath{\mathrm{u}}\xspace}

}
{

 \def\Peta        {\ensuremath{\eta}\xspace}

 \def\Ppi         {\ensuremath{\pi}\xspace}

 \mathchardef\PDelta="7101
 \mathchardef\PXi="7104
 \mathchardef\PLambda="7103
 \mathchardef\PSigma="7106
 \mathchardef\POmega="710A
 \mathchardef\PUpsilon="7107
 
 \def\PB      {\ensuremath{B}\xspace}
 
 \def\PD      {\ensuremath{D}\xspace}

 \def\PK      {\ensuremath{K}\xspace}

 \def\Pb      {\ensuremath{b}\xspace}
 \def\Pc      {\ensuremath{c}\xspace}

 \def\Pi      {\ensuremath{i}\xspace}

 \def\Pp      {\ensuremath{p}\xspace}

 \def\Ps      {\ensuremath{s}\xspace}
 
 \def\Pu      {\ensuremath{u}\xspace}

}

\makeatletter
\ifcase \@ptsize \relax
  \newcommand{\miniscule}{\@setfontsize\miniscule{4}{5}}
\or
  \newcommand{\miniscule}{\@setfontsize\miniscule{5}{6}}
\or
  \newcommand{\miniscule}{\@setfontsize\miniscule{5}{6}}
\fi
\makeatother

\DeclareRobustCommand{\optbar}[1]{\shortstack{{\miniscule (\rule[.5ex]{1.25em}{.18mm})}
  \\ [-.7ex] $#1$}}

\def\uquark    {{\ensuremath{\Pu}}\xspace}
\def\uquarkbar {{\ensuremath{\overline \uquark}}\xspace}

\def\squark    {{\ensuremath{\Ps}}\xspace}

\def\cquark    {{\ensuremath{\Pc}}\xspace}
\def\cquarkbar {{\ensuremath{\overline \cquark}}\xspace}

\def\bquark    {{\ensuremath{\Pb}}\xspace}

\def\pion   {{\ensuremath{\Ppi}}\xspace}
\def\piz    {{\ensuremath{\pion^0}}\xspace}
\def\pip    {{\ensuremath{\pion^+}}\xspace}
\def\pim    {{\ensuremath{\pion^-}}\xspace}

\def\pimp   {{\ensuremath{\pion^\mp}}\xspace}

\def\kaon    {{\ensuremath{\PK}}\xspace}
  \def\Kbar    {{\kern 0.2em\overline{\kern -0.2em \PK}{}}\xspace}

\def\KorKbar {\kern 0.18em\optbar{\kern -0.18em K}{}\xspace}

\def\Kp      {{\ensuremath{\kaon^+}}\xspace}
\def\Km      {{\ensuremath{\kaon^-}}\xspace}
\def\Kpm     {{\ensuremath{\kaon^\pm}}\xspace}

\def\Kstarz  {{\ensuremath{\kaon^{*0}}}\xspace}

\newcommand{\etapr}{\ensuremath{\Peta^{\prime}}\xspace}

  \def\Dbar    {{\kern 0.2em\overline{\kern -0.2em \PD}{}}\xspace}
\def\D       {{\ensuremath{\PD}}\xspace}

\def\DorDbar {\kern 0.18em\optbar{\kern -0.18em D}{}\xspace}
\def\Dz      {{\ensuremath{\D^0}}\xspace}
\def\Dzb     {{\ensuremath{\Dbar{}^0}}\xspace}

\def\Dstar   {{\ensuremath{\D^*}}\xspace}
\def\Dstarb  {{\ensuremath{\Dbar{}^*}}\xspace}
\def\Dstarz  {{\ensuremath{\D^{*0}}}\xspace}
\def\Dstarzb {{\ensuremath{\Dbar{}^{*0}}}\xspace}
\def\DorDstar   {{\ensuremath{\D^{(*)}}}\xspace}

\def\theDstarzb{{\ensuremath{\Dbar{}^{*}(2007)^{0}}}\xspace}

\def\theDstarm{{\ensuremath{\D^{*}(2010)^{-}}}\xspace}

\def\B       {{\ensuremath{\PB}}\xspace}
\def\Bbar    {{\ensuremath{\kern 0.18em\overline{\kern -0.18em \PB}{}}}\xspace}

\def\BorBbar    {\kern 0.18em\optbar{\kern -0.18em B}{}\xspace}
\def\Bz      {{\ensuremath{\B^0}}\xspace}
\def\Bzb     {{\ensuremath{\Bbar{}^0}}\xspace}
\def\Bu      {{\ensuremath{\B^+}}\xspace}
\def\Bub     {{\ensuremath{\B^-}}\xspace}
\def\Bp      {{\ensuremath{\Bu}}\xspace}
\def\Bm      {{\ensuremath{\Bub}}\xspace}

\def\Bd      {{\ensuremath{\B^0}}\xspace}
\def\Bs      {{\ensuremath{\B^0_\squark}}\xspace}

\def\BdorBs  {{\ensuremath{\B^0_{(\squark)}}}\xspace}

\def\Bds     {{\ensuremath{\B_{(\squark)}^0}}\xspace}

\def\Y#1S{\ensuremath{\PUpsilon{(#1S)}}\xspace}

\def\proton      {{\ensuremath{\Pp}}\xspace}
\def\antiproton  {{\ensuremath{\overline \proton}}\xspace}

\def\Lz          {{\ensuremath{\PLambda}}\xspace}
\def\Lbar        {{\ensuremath{\offsetoverline{\PLambda}}}\xspace}
\def\LorLbar     {\kern 0.18em\optbar{\kern -0.18em \PLambda}{}\xspace}

\def\Lb           {{\ensuremath{\Lz^0_\bquark}}\xspace}
\def\Lbbar        {{\ensuremath{\Lbar{}^0_\bquark}}\xspace}

\newcommand{\decay}[2]{\mbox{\ensuremath{#1\!\to #2}}\xspace}         

\def\to                 {\ensuremath{\rightarrow}\xspace}

\def\AT#1     {\ensuremath{A_{\mathrm{T}}^{#1}}\xspace}           

\def\C#1      {\ensuremath{\mathcal{C}_{#1}}\xspace}                       
\def\Cp#1     {\ensuremath{\mathcal{C}_{#1}^{'}}\xspace}                    
\def\Ceff#1   {\ensuremath{\mathcal{C}_{#1}^{\mathrm{(eff)}}}\xspace}        
\def\Cpeff#1  {\ensuremath{\mathcal{C}_{#1}^{'\mathrm{(eff)}}}\xspace}       
\def\Ope#1    {\ensuremath{\mathcal{O}_{#1}}\xspace}                       
\def\Opep#1   {\ensuremath{\mathcal{O}_{#1}^{'}}\xspace}                    

\newcommand{\nospaceunit}[1]{\ensuremath{\text{#1}}}
\newcommand{\aunit}[1]{\ensuremath{\text{\,#1}}}

\newcommand{\tev}{\aunit{Te\kern -0.1em V}\xspace}
\newcommand{\gev}{\aunit{Ge\kern -0.1em V}\xspace}
\newcommand{\mev}{\aunit{Me\kern -0.1em V}\xspace}
\newcommand{\kev}{\aunit{ke\kern -0.1em V}\xspace}
\newcommand{\ev}{\aunit{e\kern -0.1em V}\xspace}
\newcommand{\mevc}{\ensuremath{\aunit{Me\kern -0.1em V\!/}c}\xspace}
\newcommand{\gevc}{\ensuremath{\aunit{Ge\kern -0.1em V\!/}c}\xspace}
\newcommand{\mevcc}{\ensuremath{\aunit{Me\kern -0.1em V\!/}c^2}\xspace}
\newcommand{\gevcc}{\ensuremath{\aunit{Ge\kern -0.1em V\!/}c^2}\xspace}

\def\mum  {\ensuremath{\,\upmu\nospaceunit{m}}\xspace}

\def\fb   {\ensuremath{\aunit{fb}}\xspace}
\def\invfb   {\ensuremath{\fb^{-1}}\xspace}

\def\ps   {\ensuremath{\aunit{ps}}\xspace}

\newcommand{\chisq}{\ensuremath{\chi^2}\xspace}

\newcommand{\chisqip}{\ensuremath{\chi^2_{\text{IP}}}\xspace}

\def\gsim{{~\raise.15em\hbox{$>$}\kern-.85em
          \lower.35em\hbox{$\sim$}~}\xspace}
\def\lsim{{~\raise.15em\hbox{$<$}\kern-.85em
          \lower.35em\hbox{$\sim$}~}\xspace}

\def\sPlot{\mbox{\em sPlot}\xspace}

\def\pt         {\ensuremath{p_{\mathrm{T}}}\xspace}

\def\ptot       {\ensuremath{p}\xspace}

\def\degrees{\ensuremath{^{\circ}}\xspace}

\def\evtgen     {\mbox{\textsc{EvtGen}}\xspace}

\def\geant      {\mbox{\textsc{Geant4}}\xspace}

\def\photos     {\mbox{\textsc{Photos}}\xspace}

\def\pythia     {\mbox{\textsc{Pythia}}\xspace}

\xspace

\def\tell1  {TELL1\xspace}
\def\ukl1   {UKL1\xspace}

\newcommand{\eg}{\mbox{\itshape e.g.}\xspace}
\newcommand{\ie}{\mbox{\itshape i.e.}\xspace}

\newcommand{\etc}{\mbox{\itshape etc.}\xspace}

\usepackage{cite} 
\usepackage{mciteplus}

\usepackage{tikz-feynman} 
\usepackage{multirow}
\usepackage[capitalise]{cleveref}

\begin{document}

\renewcommand{\thefootnote}{\fnsymbol{footnote}}
\setcounter{footnote}{1}
\begin{titlepage}
\pagenumbering{roman}

\vspace*{-1.5cm}
\centerline{\large EUROPEAN ORGANIZATION FOR NUCLEAR RESEARCH (CERN)}
\vspace*{1.5cm}
\noindent
\begin{tabular*}{\linewidth}{lc@{\extracolsep{\fill}}r@{\extracolsep{0pt}}}
\ifthenelse{\boolean{pdflatex}}
{\vspace*{-1.5cm}\mbox{\!\!\!\includegraphics[width=.14\textwidth]{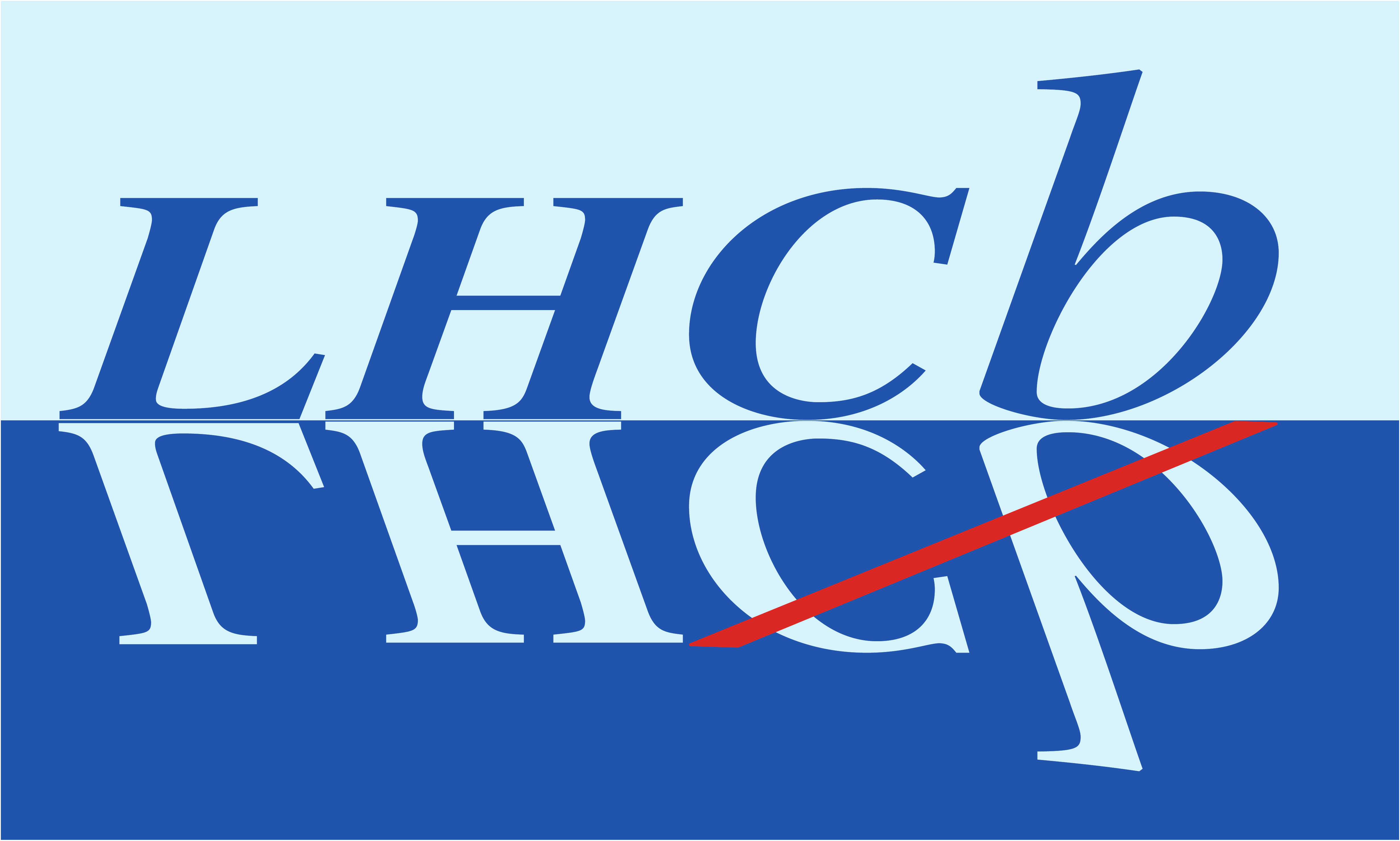}} & &}%
{\vspace*{-1.2cm}\mbox{\!\!\!\includegraphics[width=.12\textwidth]{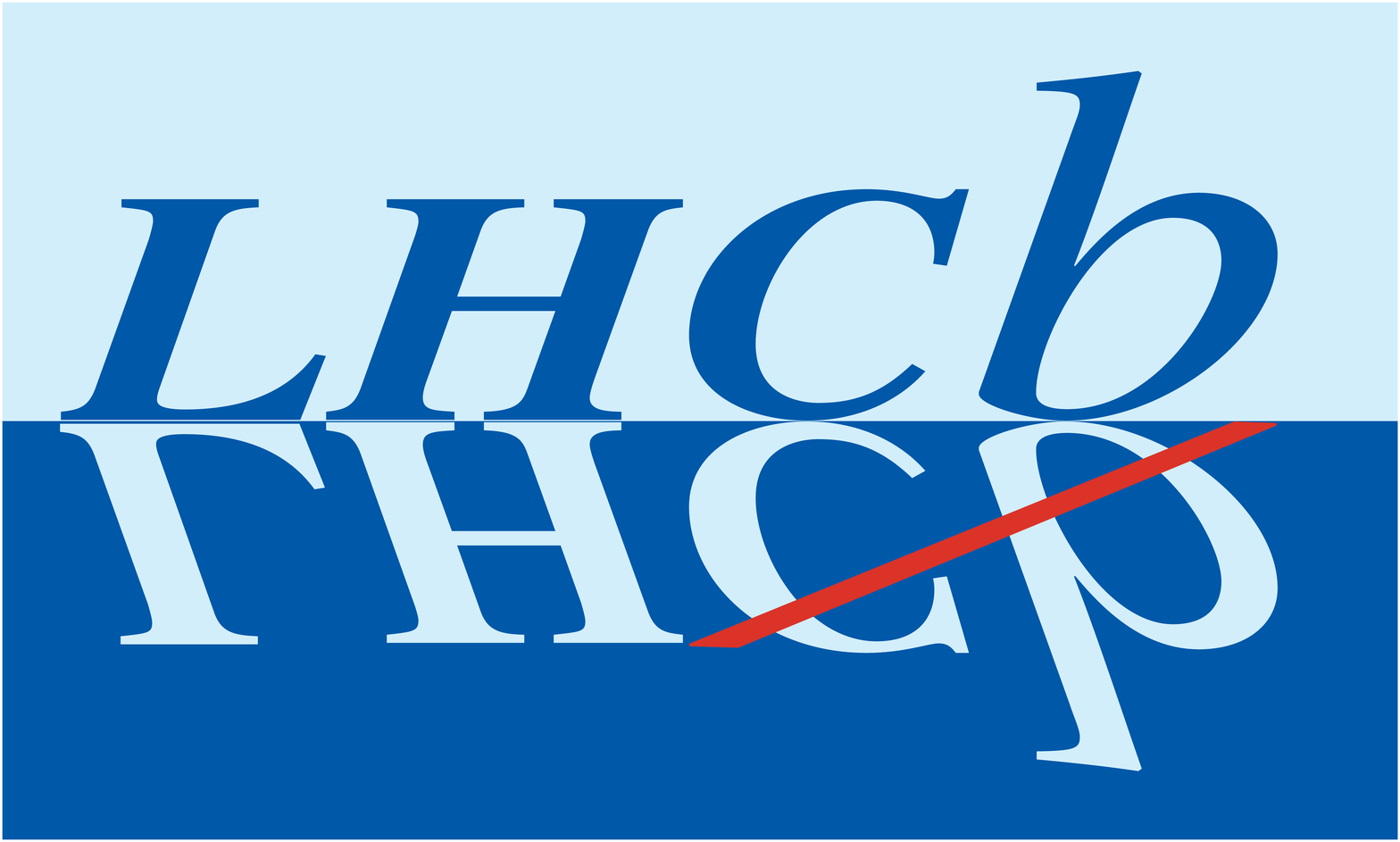}} & &}%
\\
 & & CERN-EP-2021-262 \\  
 & & LHCb-PAPER-2021-043 \\  
 & & May 12, 2022 \\ 
 & & \\
\end{tabular*}

\vspace*{4.0cm}

{\normalfont\bfseries\boldmath\huge
\begin{center}
  \papertitle 
\end{center}
}

\vspace*{2.0cm}

\begin{center}
\paperauthors\footnote{Authors are listed at the end of this paper.}
\end{center}

\vspace{\fill}

\begin{abstract}
  \noindent
  The first observations of $\Bz\to\theDstarzb\Kp\pim$ and $\Bs\to\theDstarzb\Km\pip$ decays are presented, and their branching fractions relative to that of the \mbox{$\Bz\to\theDstarzb\pip\pim$} decay are reported.
  These modes can potentially be used to investigate the spectroscopy of charm and charm-strange resonances and to determine the angle $\gamma$ of the CKM unitarity triangle.
  It is also important to understand them as a source of potential background in determinations of $\gamma$ from $\Bp \to D\Kp$ and $\Bz \to D\Kp\pim$ decays.
  The analysis is based on a sample corresponding to an integrated luminosity of $5.4 \invfb$ of proton--proton collision data at $13\tev$ centre-of-mass energy recorded with the LHCb detector.
  The $\theDstarzb$ mesons are fully reconstructed in the $\Dzb\piz$ and $\Dzb\gamma$ channels, with the $\Dzb \to \Kp\pim$ decay.
  A novel weighting method is used to subtract background while simultaneously applying an event-by-event efficiency correction to account for resonant structures in the decays.
\end{abstract}

\vspace*{2.0cm}

\begin{center}
  Published in Phys.~Rev.~D 105, 072005 (2022)
\end{center}

\vspace{\fill}

{\footnotesize 
  \centerline{\copyright~\papercopyright. \href{\paperlicenceurl}{\paperlicence}.}}
\vspace*{2mm}

\end{titlepage}


\newpage
\setcounter{page}{2}
\mbox{~}

\renewcommand{\thefootnote}{\arabic{footnote}}
\setcounter{footnote}{0}


\pagestyle{plain} 
\setcounter{page}{1}
\pagenumbering{arabic}


\section{Introduction}
\label{sec:introduction}

The precise measurement of the angle $\gamma$ of the Cabibbo-Kobayashi-Maskawa (CKM) unitarity triangle is among the priorities of current particle physics experiments.
It can be determined by exploiting the interference between the \decay{\bquark}{\cquark\uquarkbar\squark} and \decay{\bquark}{\uquark\cquarkbar\squark} amplitudes in many \B to open-charm decays.
In this approach the determination of $\gamma$ has a tiny theoretical uncertainty in the Standard Model~\cite{Brod:2013sga}, since both amplitudes correspond to tree-level processes and all hadronic parameters that quantify the extent of the interference can be determined from data.

The world average value of $\gamma$ currently has an uncertainty of around $4\degrees$~\cite{HFLAV18,PDG2020,LHCb-PAPER-2021-033}.
This level of precision can be significantly improved not only by increasing the size of the data samples, but also by studying new decay channels.
One such channel, which has not yet been exploited fully, is the $\Bd \to \D\Kp\pim$ decay where the symbol $\D$ is used to denote a neutral \D\ meson that is any admixture of the $\Dz$ and $\Dzb$ states.\footnote{
    The inclusion of charge conjugate processes is implied throughout the document.}
A Dalitz-plot analysis of the $\Bd \to \D\Kp\pim$ decay provides additional sensitivity to $\gamma$ compared to a quasi-two-body analysis of the $\Bd \to \D\Kstarz$ channel~\cite{Gershon:2008pe,Gershon:2009qc}.
Such a Dalitz-plot analysis has been performed by the LHCb collaboration~\cite{LHCb-PAPER-2015-059}, however the results suffer from significant systematic uncertainties due to background from \decay{\Bds}{\Dstarzb\Kpm\pimp} decays. 
The symbol $\Dstarzb$ is used throughout this paper to denote the $\theDstarzb$ meson.
In particular, there is a large background from \decay{\Bs}{\Dstarzb\Km\pip} decays, where the soft neutral particle from $\Dstarzb\to\Dzb\piz$ or $\Dzb\gamma$ process is not reconstructed, followed by the favoured $\Dzb\to\Kp\pim$ transition, when attempting to reconstruct the suppressed $\Bzb\to D\Km\pip$, $D \to \Kp\pim$ channel.
Similar background is also seen to contribute to the selected sample in analyses of \decay{\Bm}{\DorDstar\Km} decays~\cite{LHCb-PAPER-2020-019,LHCb-PAPER-2020-036}.
Improved understanding of the branching fractions and Dalitz plots of \decay{\Bds}{\Dstarzb\Kpm\pimp} decays is therefore important to control systematic uncertainties in future analyses and achieve precise constraints on $\gamma$.

The \decay{\Bd}{\Dstar\Kp\pim} decay, where \Dstar\ denotes any admixture of \Dstarz\ and \Dstarzb\ mesons, could itself potentially be used to gain additional sensitivity to $\gamma$.
Measurements of the properties of this decay are essential to understand the potential gains.
In addition, studies of the Dalitz plots of both \decay{\Bs}{\Dstarzb\Km\pip} and \decay{\Bd}{\Dstarzb\Kp\pim} decays provide opportunities to investigate, respectively, charm-strange and charm meson spectroscopy. 
Investigations of $B_{(s)} \to \Dbar hh^\prime$ decays, where $h^{(\prime)}$ is a pion or kaon, have provided a wealth of information in this field~\cite{Belle:2003nsh,Belle:2006wbx,BaBar:2009pnd,LHCb-PAPER-2014-035,LHCb-PAPER-2014-036,LHCb-PAPER-2015-007,LHCb-PAPER-2014-070,LHCb-PAPER-2015-017,LHCb-PAPER-2016-026}, but there are fewer studies to date of the corresponding $B_{(s)} \to \Dstarb hh^\prime$ decays~\cite{Belle:2003nsh,LHCb-PAPER-2017-006,LHCb-PAPER-2019-027}.
Since resonances with unnatural spin-parity quantum numbers ($J^P = 0^-, 1^+, 2^-$, \etc) cannot decay to two pseudoscalar particles, studies of the $\Dstarb h$ spectra could reveal structures that cannot be seen in the $\Dbar h$ final state.  

\begin{figure}[!tb]
\centering
	\includegraphics[width=0.85\textwidth]{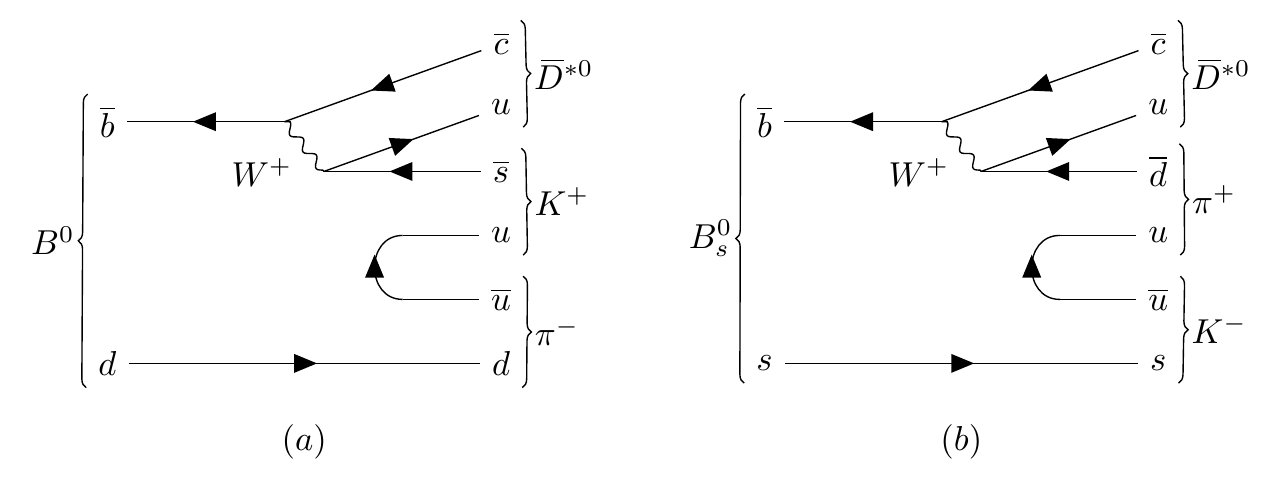}
  	\includegraphics[width=0.85\textwidth]{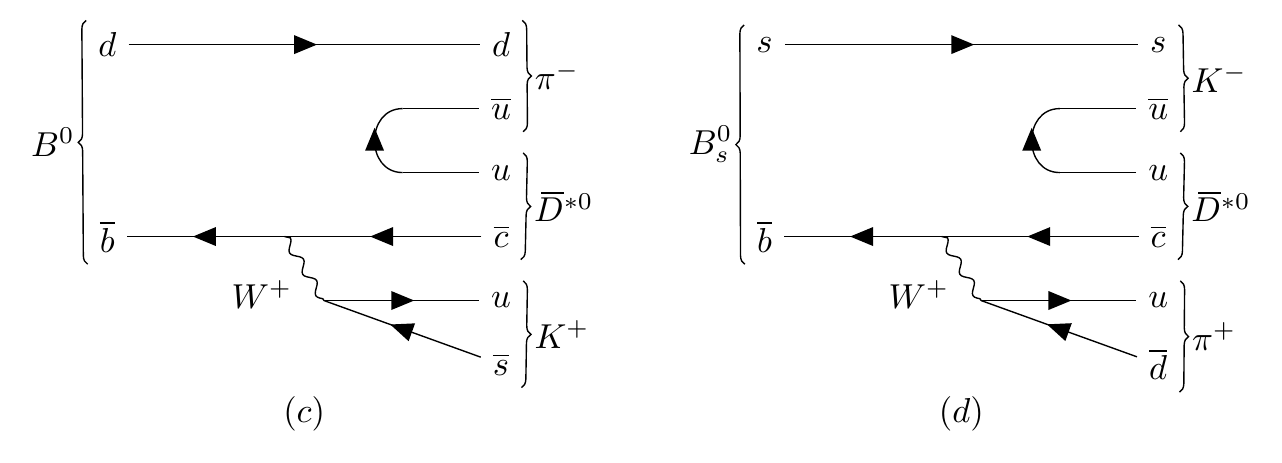}
\caption{
    Leading diagrams for ($a$), ($c$)~\decay{\Bz}{\Dstarzb\Kp\pim} and ($b$), ($d$)~\decay{\Bs}{\Dstarzb\Km\pip} decay channels. 
    In the colour-suppressed diagrams, ($a$) and ($b$), the decay proceeds through an intermediate $\KorKbar{}^{*0}$ resonance with additional $\uquark\uquarkbar$ production, while in the colour-allowed diagrams the decay proceeds through an excited charm~($c$) or charm-strange~($d$) resonance.
    }
\label{fig:feynman-diagrams}
\end{figure}

Tree-level diagrams for \decay{\Bd}{\Dstarzb\Kp\pim} and \decay{\Bs}{\Dstarzb\Km\pip} decays are shown in \cref{fig:feynman-diagrams}, indicating how various resonances can be produced.
The resonant substructures of these decays, together with significant variation of the signal efficiency across the phase space, make it necessary to use an event-by-event efficiency correction in order to avoid a limiting systematic uncertainty when measuring the branching fractions.
A novel approach to background subtraction is employed in order to achieve this in a robust way.
While this approach ensures that the Dalitz plot of each decay mode is accounted for in the efficiency evaluation, a full amplitude analysis is out of the scope of this study.

There are no previous experimental results on the \decay{\Bs}{\Dstarzb\Km\pip} and \decay{\Bd}{\Dstarzb\Kp\pim} decay channels.
The branching fraction of \decay{\Bd}{\Dstarzb\pip\pim} decays has been measured by the Belle collaboration~\cite{Satpathy:2002js} to be ${\cal B}(\decay{\Bd}{\Dstarzb\pip\pim}) = (6.2 \pm 1.2 \pm 1.8) \times 10^{-4}$, where the first uncertainty is statistical and the second is systematic. 
Therefore this mode is used for normalisation of the branching fraction measurements.
Both \decay{\Dstarzb}{\Dzb\gamma} and \decay{\Dstarzb}{\Dzb\piz} decays are fully reconstructed, and in all cases the favoured $\Dzb \to \Kp\pim$ decay is used.
Separate measurements of the branching fraction ratios are made for the two \Dstarzb\ decays, in order to reduce potential systematic uncertainties, and the results are combined accounting for correlations.
The analysis is based on a data sample corresponding to an integrated luminosity of $5.4 \invfb$ of LHC proton–proton ($pp$) collisions, at centre-of-mass energy of $13\tev$, recorded by the LHCb detector during the years 2016, 2017 and 2018.

The remainder of the paper is organised as follows.
The LHCb detector and software is described in Sec.~\ref{sec:detector}.
The selection of signal candidates is discussed in Sec.~\ref{sec:selection}, with the procedure to determine the yields through a simultaneous fit to their mass distributions outlined in Sec.~\ref{sec:massfits}.
The event-by-event weighting and efficiency-correction procedure is described in Sec.~\ref{sec:efficiency}.
The systematic uncertainties are detailed in Sec.~\ref{sec:systematics}, with the results of the analysis discussed in Sec.~\ref{sec:results}.

\section{LHCb detector and software}
\label{sec:detector}

The \lhcb detector~\cite{LHCb-DP-2008-001,LHCb-DP-2014-002} is a single-arm forward spectrometer covering the \mbox{pseudorapidity} range $2<\eta <5$, designed for the study of particles containing \bquark or \cquark quarks.
The detector includes a high-precision tracking system consisting of a silicon-strip vertex detector surrounding the $pp$ interaction region~\cite{LHCb-DP-2014-001}, a large-area silicon-strip detector located upstream of a dipole magnet with a bending power of about $4{\mathrm{\,Tm}}$, and three stations of silicon-strip detectors and straw drift tubes~\cite{LHCb-DP-2017-001} placed downstream of the magnet.
The tracking system provides a measurement of the momentum, \ptot, of charged particles with relative uncertainty that varies from 0.5\% at low momentum to 1.0\% at 200\gevc.
The minimum distance of a track to a primary $pp$ collision vertex (PV), the impact parameter (IP), is measured with resolution of $(15+29/\pt)\mum$, where \pt is the component of the momentum transverse to the beam, in~\gevc.
Different species of charged hadrons are distinguished using information from two ring-imaging Cherenkov detectors~\cite{LHCb-DP-2012-003}.
Photons, electrons and hadrons are identified by a calorimeter system consisting of scintillating-pad and preshower detectors, an electromagnetic and a hadronic calorimeter~\cite{LHCb-DP-2020-001}.
The resolution of the energy measurement in the electromagnetic calorimeter varies from around 12\% at $3\gev$ to around $5\%$ at $80 \gev$.
Muons are identified by a system composed of alternating layers of iron and multiwire proportional chambers~\cite{LHCb-DP-2012-002}.
The online event selection is performed by a trigger~\cite{LHCb-DP-2012-004}, which consists of a hardware stage, based on information from the calorimeter and muon systems, followed by a software stage, which applies a full event reconstruction.
At the hardware trigger stage, events are required to have a muon with high \pt or a hadron, photon or electron with high transverse energy in the calorimeters.
For hadrons, the transverse energy threshold is $3.5\gev$.
The software trigger requires a two-, three- or four-track secondary vertex with significant displacement from any primary $pp$ interaction vertex.
At least one charged particle must have significant transverse momentum and be inconsistent with originating from a PV.
A multivariate algorithm~\cite{BBDT,LHCb-PROC-2015-018} is used for the identification of secondary vertices consistent with the decay of a \bquark hadron.

Simulation is used to model the effects of the detector acceptance and response, and the imposed selection requirements.
In the simulation, $pp$ collisions are generated using \pythia~\cite{Sjostrand:2007gs,*Sjostrand:2006za} with a specific \lhcb configuration~\cite{LHCb-PROC-2010-056}.
Decays of unstable particles are described by \evtgen~\cite{Lange:2001uf}, in which final-state radiation is generated using \photos~\cite{davidson2015photos}.
The interaction of the generated particles with the detector, and its response, are implemented using the \geant toolkit~\cite{Allison:2006ve, *Agostinelli:2002hh} as described in Ref.~\cite{LHCb-PROC-2011-006}. 
The underlying $pp$ interaction is reused multiple times, with an independently generated signal decay for each~\cite{LHCb-DP-2018-004}.

\section{Selection of signal candidates}
\label{sec:selection}

The selection procedures follow those used in previous LHCb analyses of $\BdorBs \to \D\Kpm\pimp$ decays~\cite{LHCb-PAPER-2014-035,LHCb-PAPER-2014-036,LHCb-PAPER-2015-017}, with modifications to account for the selection of \Dstarzb\ mesons. 
Selection requirements are imposed in an initial filtering stage, with multivariate analysis techniques subsequently used to separate signal from background.
The variables used are related to the decay kinematics and topology, exploiting the capability of the LHCb detector to reconstruct precisely particle momenta as well as production and decay vertices. 
Since the final requirements imposed are tighter than those in the filtering stage, details of most of the initial selections are omitted here for brevity.  
The requirements are common for all six channels --- $\Bd \to \Dstarzb\Kp\pim$, $\Bs \to \Dstarzb\Km\pip$ and $\Bd \to \Dstarzb\pip\pim$, each with both $\Dstarzb \to \Dzb\gamma$ and $\Dzb\piz$ decays --- except for charged hadron identification requirements and $\Dstarzb$ reconstruction.

Candidate \BdorBs\ decays are composed of four final-state charged particles and the soft neutral products ($\gamma$ or $\piz$) of the \Dstarzb\ decay.
All charged and neutral final-state particles are required to have $p$ and \pt\ values above thresholds designed to reduce the large background of soft random particles originating from the rest of the event.
The tracks corresponding to the charged particles are required to be of good quality and to be significantly displaced from any PV as quantified through the \chisqip\ variable, where \chisqip\ is defined as the difference in the vertex-fit \chisq\ of a given PV reconstructed with and without the particle under consideration.
A \Dzb\ candidate is formed from two oppositely charged tracks, assigned the kaon and pion mass hypotheses, with invariant mass close to the known \Dzb\ mass~\cite{PDG2020}.
These two tracks must be consistent with originating from a common vertex that is significantly displaced from any PV.
Photon (neutral pion) candidates are formed from a single (two resolved) clusters of deposited energy in the calorimeter system~\cite{LHCb-DP-2020-001}, and are combined with the \Dzb\ candidate to form a \Dstarzb\ candidate.

Trigger signals are associated with the particles reconstructed offline.
It is required that the hardware-stage triggering of the event containing a candidate is either due to an energy deposit in the calorimeters associated with the signal, or due to other particles produced in the $pp$ collision without the involvement of signal particles.

\begin{figure}[!tb]
  \centering
  \includegraphics[width=0.45\textwidth]{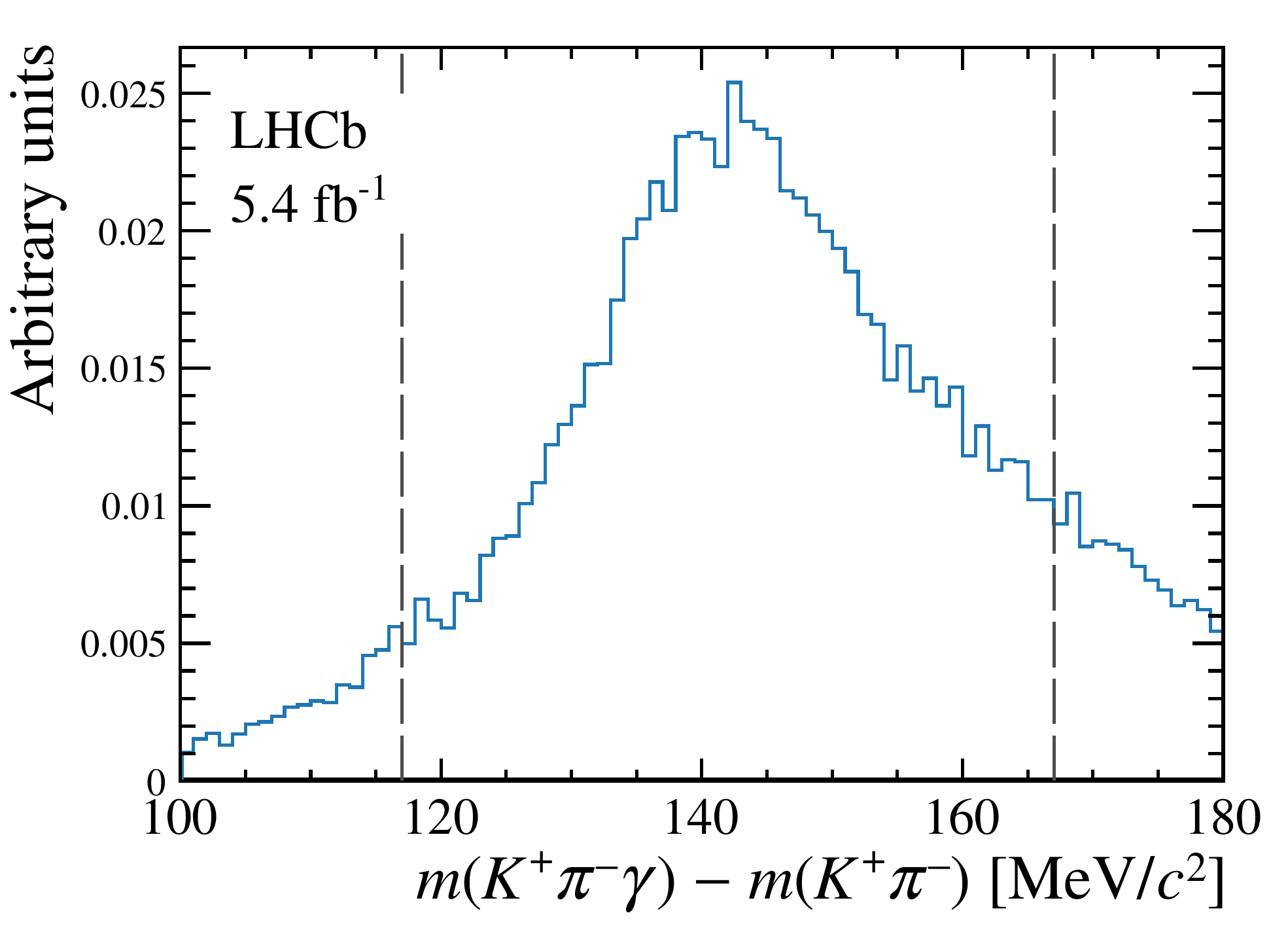}
  \includegraphics[width=0.45\textwidth]{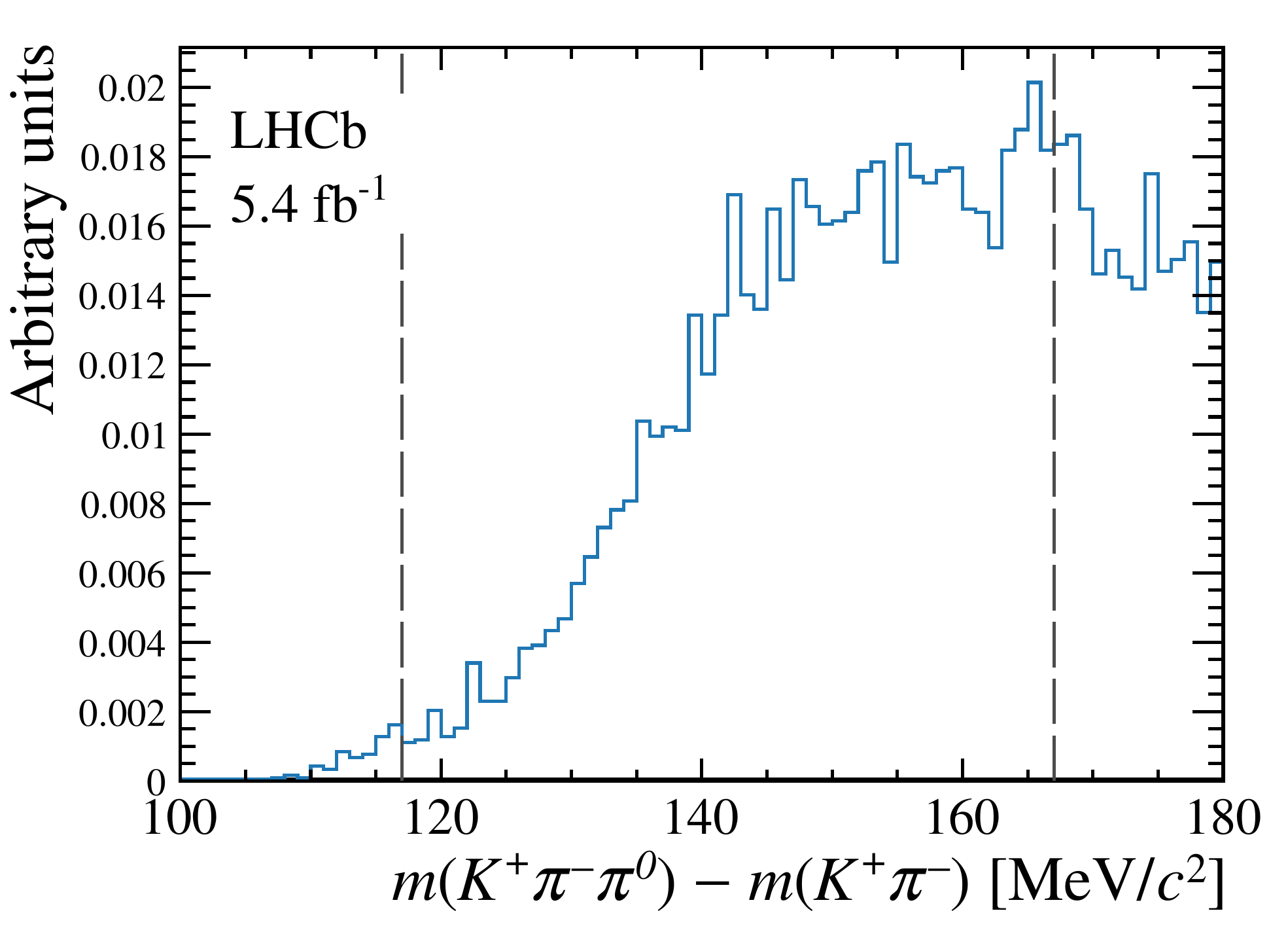}
  \includegraphics[width=0.45\textwidth]{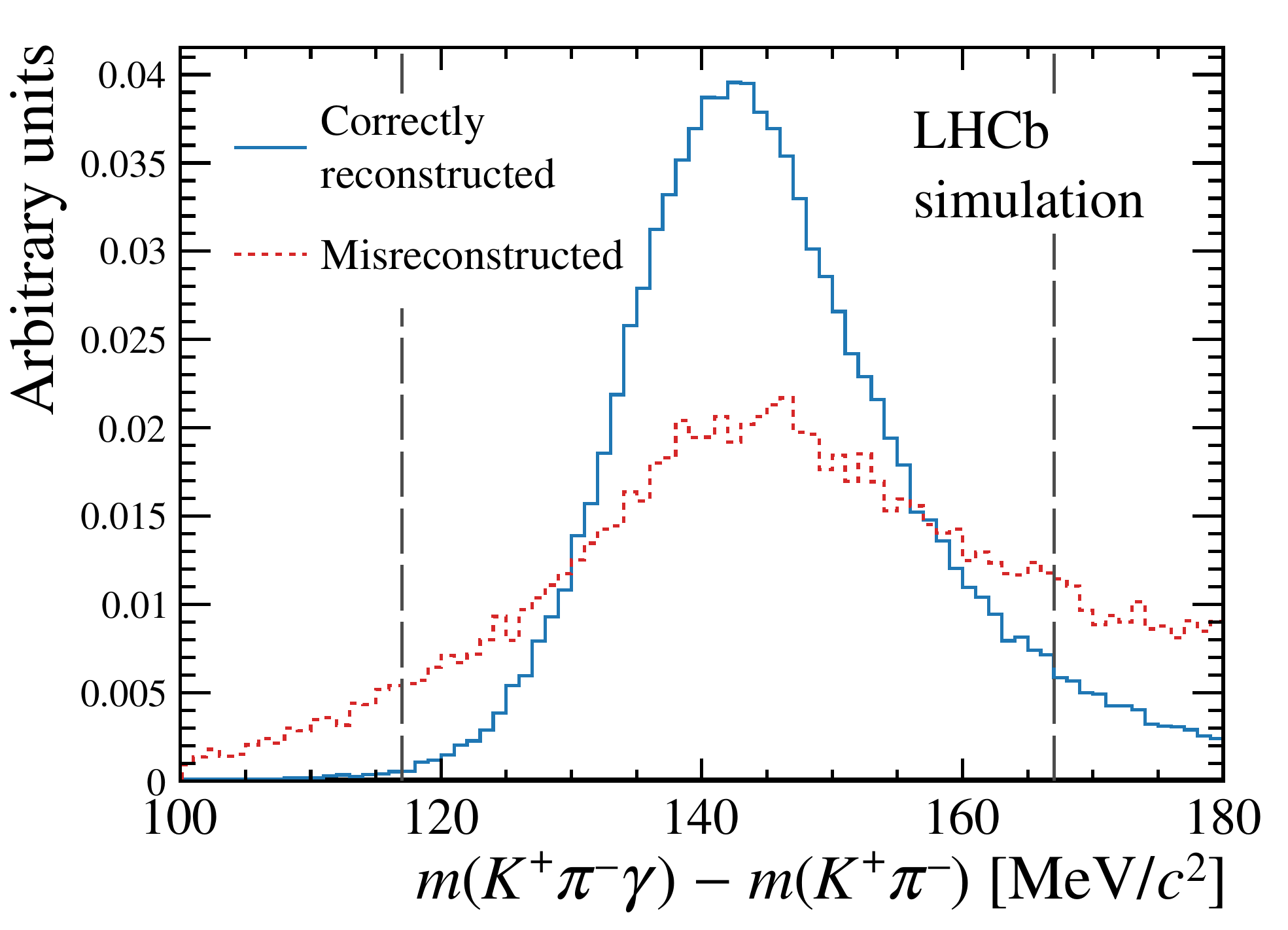}
  \includegraphics[width=0.45\textwidth]{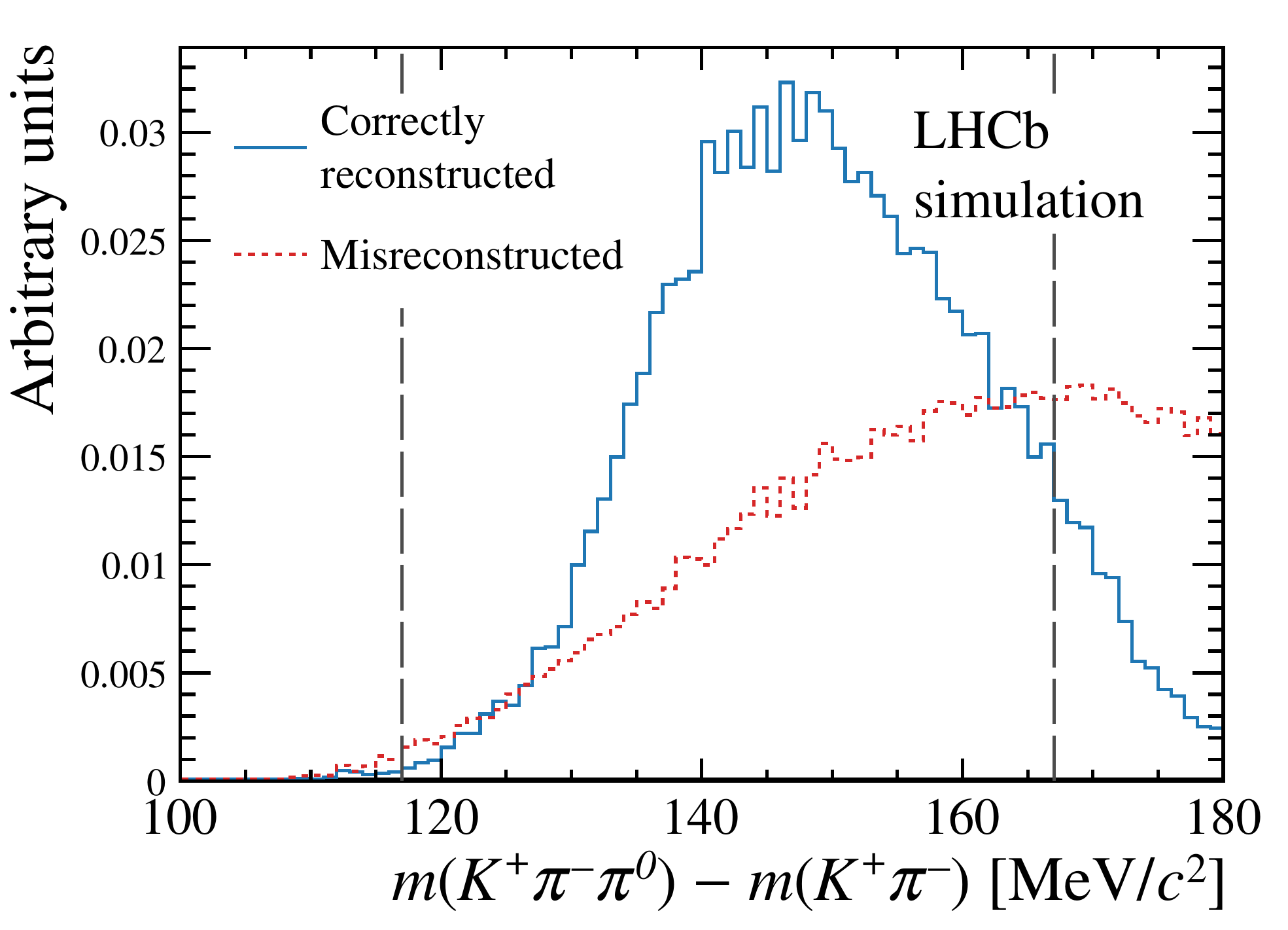}
  \caption{
    Difference between the \Dstarzb\ and \Dzb\ candidate masses in $\Bd\to\Dstarzb\pip\pim$ decays for (top)~data and (bottom)~simulated signal decays.
    The left and right plots show $\Dstarzb\to\Dzb\gamma$ and $\Dzb\piz$ candidates, respectively.
    The simulation distributions are shown separately for correctly reconstructed and misreconstructed signal components.
    All selection requirements have been imposed, except those on the plotted variable, which are indicated with dashed vertical lines.
  }
  \label{fig:Dstarmass}
\end{figure}

The difference between the \Dstarzb\ and \Dzb\ candidate masses is required to be in the range $117\text{--}167 \mevcc$, encompassing the known value $142 \mevcc$~\cite{PDG2020}, for both \Dstarzb\ decay modes. 
The distribution of this variable in data and in simulated signal decays is shown in \cref{fig:Dstarmass}.
Although around $25\%$ of the signal is lost due to this requirement, it is necessary to reduce the otherwise potentially overwhelming background.

The \Dstarzb\ candidate is combined with the other two charged particles to form a \BdorBs\ candidate, again with requirements on invariant mass, vertex consistency and displacement from each PV.
The primary vertex that fits best to the flight direction of the \BdorBs\ candidate is taken as the associated PV.
A requirement is imposed on the \chisq\ of a kinematic fit~\cite{Hulsbergen:2005pu} to the candidate’s production and decay chain, in which the \Dzb\ and \Dstarzb\ masses are constrained to their known values~\cite{PDG2020}.

The signature of a charm hadron originating from a $B$ decay (referred to as $D$ from $B$) provides discrimination power against background that either does not involve a real charm decay or where the charm hadron originates promptly from the PV.
A Neural Network (NN) algorithm~\cite{NeuroBayes} is trained to identify this signature, using $\Bp\rightarrow \Dzb\pip, \Dzb\rightarrow \Kp\pim$ decays in data, with the same preselection requirements as for the \BdorBs\ candidates.
Signal and background are separated using the \sPlot\ procedure~\cite{Pivk:2004ty}, so that the training is entirely data-driven.
The input variables are the $p$, \pt, \chisqip and track quality of the \Dzb\ decay products together with the $p$, \pt, \chisqip, vertex quality and flight distance of the \Dzb\ candidate.
An initial loose requirement is imposed on the output of the $D$ from $B$ NN classifier, and this variable is also used in the subsequent multivariate analysis.

Two boosted decision tree~(BDT) algorithms~\cite{Breiman,AdaBoost,XGBoost} are used to separate further signal from background.
The first is designed to reduce background where the final-state particles are misidentified.
Such background may come from other $B$ decays --- including cross-feed between signal modes --- or from random combinations of particles, subsequently referred to as combinatorial background.
Separate classifiers are trained for $\BdorBs \to \Dstarzb\Kpm\pimp$ and $\Bd \to \Dstarzb\pip\pim$ decays, with background samples comprised of a cocktail of $\BdorBs$ decays with misidentified final-state particles. 
All training samples are taken from simulation.
This BDT classifier takes as input variables derived from the output of the ring-imaging Cherenkov detectors, which distinguish charged pions and kaons.
In order to ensure good agreement with the distributions in data, the values of these variables in the simulation are sampled from data control samples with a procedure that accounts for correlations between the variables associated to a particular track, as well as the dependence of the response on track kinematics~\cite{LHCb-DP-2018-001}.
The BDT algorithm also exploits the $D$ from $B$ NN output, as well as kinematic and topological variables related to the \BdorBs\ and \Dzb\ decays, since correlations between these and the charged hadron identification variables help to improve the separation power.  

The second BDT classifier is designed to reduce combinatorial background, and is trained separately for the \Dstarzb\ decays to $\Dzb\gamma$ and $\Dzb\piz$.
It takes as input kinematic and topological variables, together with the $D$ from $B$ NN output, quantities related to the confidence of correctly identifying a photon or neutral pion from the calorimeter information, and variables that quantify how well the signal \BdorBs\ candidate is isolated from other activity in the $pp$ collision event~\cite{LHCb-PAPER-2014-036}.
The second BDT algorithm is trained using a signal sample taken from simulation and a background sample taken from the data sidebands with \BdorBs-candidate mass in the range $5000\text{--}5150 \mevcc$ or $5450\text{--}6000 \mevcc$.

Requirements on the two BDT outputs are chosen simultaneously by optimising the figure-of-merit ${\cal S}/\sqrt{{\cal S}+{\cal B}}$, where ${\cal S}$ and ${\cal B}$ are the expected signal and background yields in the signal region of $\pm50\mevcc$ around the known $\BdorBs$ mass~\cite{PDG2020}.  
The reference value for ${\cal S}$ is obtained from a fit to the \BdorBs-candidate mass distributions with tight cuts applied, and that for ${\cal B}$ is obtained from a simple fit with no requirement on the BDT outputs.
The value of ${\cal S}$ at different BDT output requirements is extrapolated from the reference value using efficiencies from simulation, while ${\cal B}$ is extrapolated according to the number of candidates retained in the high-mass sideband.

The optimisation procedure results in very similar requirements for channels with similar final states, and to minimise systematic uncertainties the same requirements are imposed.  
Specifically, the $\Bd \to \Dstarzb\Kp\pim$ and $\Bs \to \Dstarzb\Km\pip$ candidates share the same requirement on the first BDT classifier, while all modes with $\Dstarzb \to \Dzb\gamma$ decays share the same requirement on the second BDT algorithm, as do all modes with $\Dstarzb \to \Dzb\piz$ decays.
Relative to the sample after all preselection requirements, the BDT requirements have combined efficiencies of around 50\%~(60\%) for $\BdorBs \to \Dstarzb\Kpm\pimp$ decays with $\Dstarzb \to \Dzb\gamma$ ($\Dstarzb \to \Dzb\piz$), while the efficiencies for $\Bd \to \Dstarzb\pip\pim$ decays are higher due to the requirement on the first BDT classifier being looser.
The requirements reduce the background levels by around two~(one) orders of magnitude in the $\BdorBs \to \Dstarzb\Kpm\pimp$ ($\Bd \to \Dstarzb\pip\pim$) samples. 
Finally, only candidates with \BdorBs-candidate mass in the range $5100\text{--}5900\mevcc$ are retained for the fit to determine the signal yields.

After all selection requirements are imposed, there are a small number of cases where candidates from the same $pp$ collision event are seen in two different final states (referred to subsequently as duplicate candidates).
In particular, as the selection requirements for the two \Dstarzb\ decays are not mutually exclusive, around $5\%$ of events with candidates in channels involving $\Dstarzb \to \Dzb\gamma$ decays also contain candidates in the same final state except with $\Dstarzb \to \Dzb\piz$.
Similarly, a single event can produce multiple candidates in the same final state, which happens at a rate of around $4\%$ ($9\%$) of selected candidates in channels with $\Dstarzb \to \Dzb\gamma$ ($\Dstarzb \to \Dzb\piz$) decays.  
These rates are similar to those seen in simulation.
All such candidates are retained, with their presence considered as a source of systematic uncertainty in the results.

\section{Determination of signal yields}
\label{sec:massfits}

\begin{figure}[!tb]
  \centering
  \includegraphics[width=0.48\linewidth]{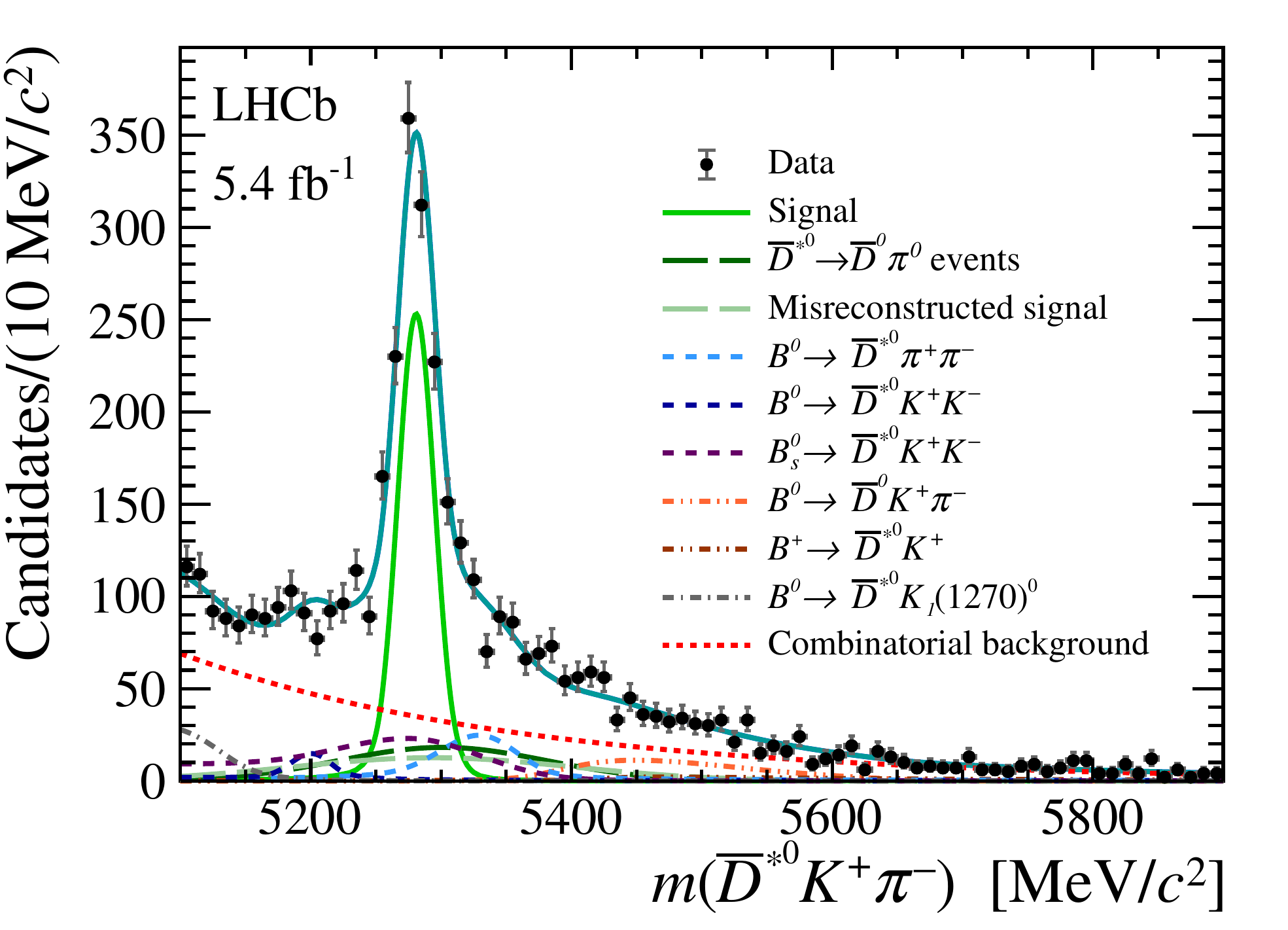}
  \includegraphics[width=0.48\linewidth]{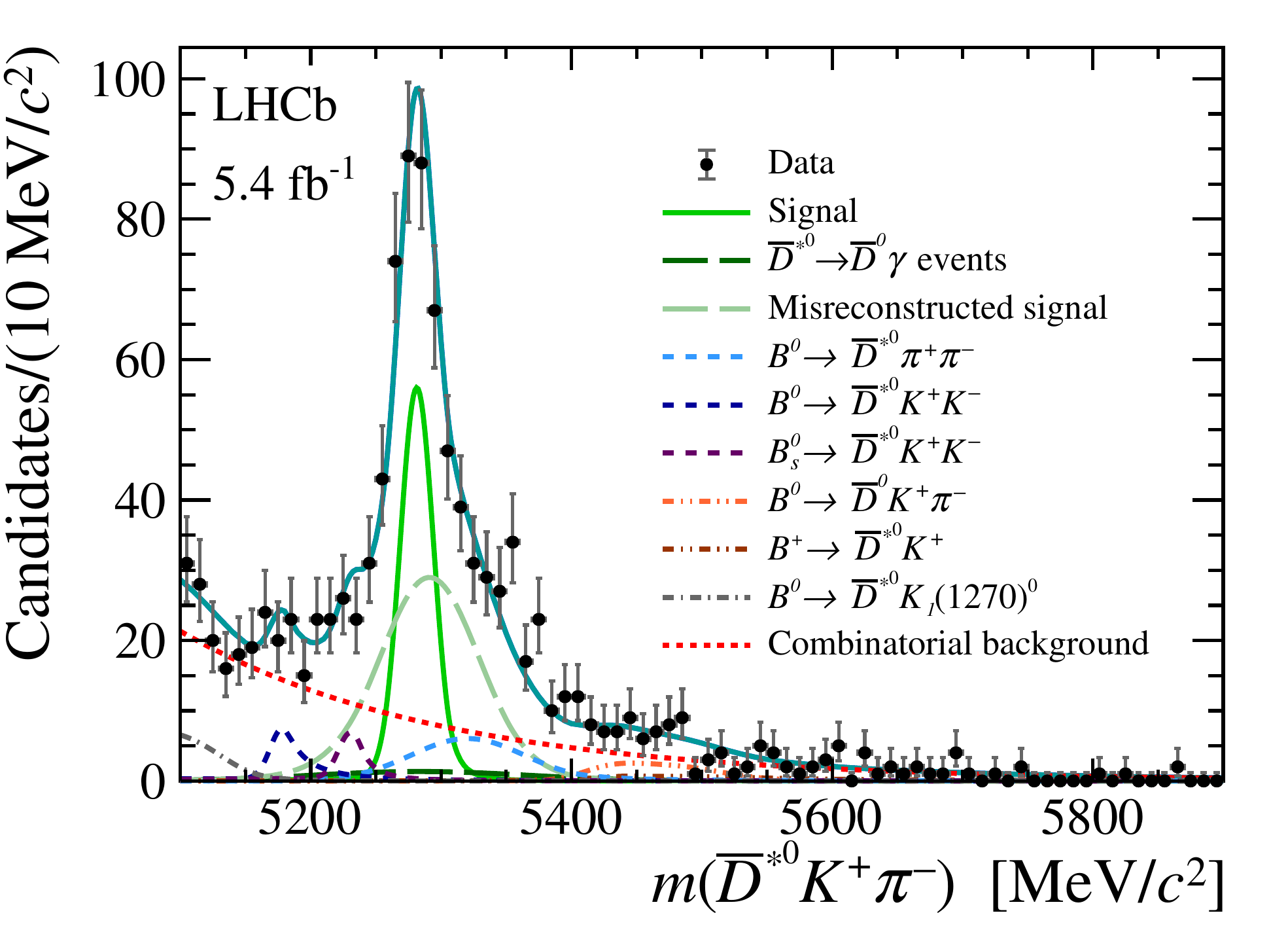}
  \includegraphics[width=0.48\linewidth]{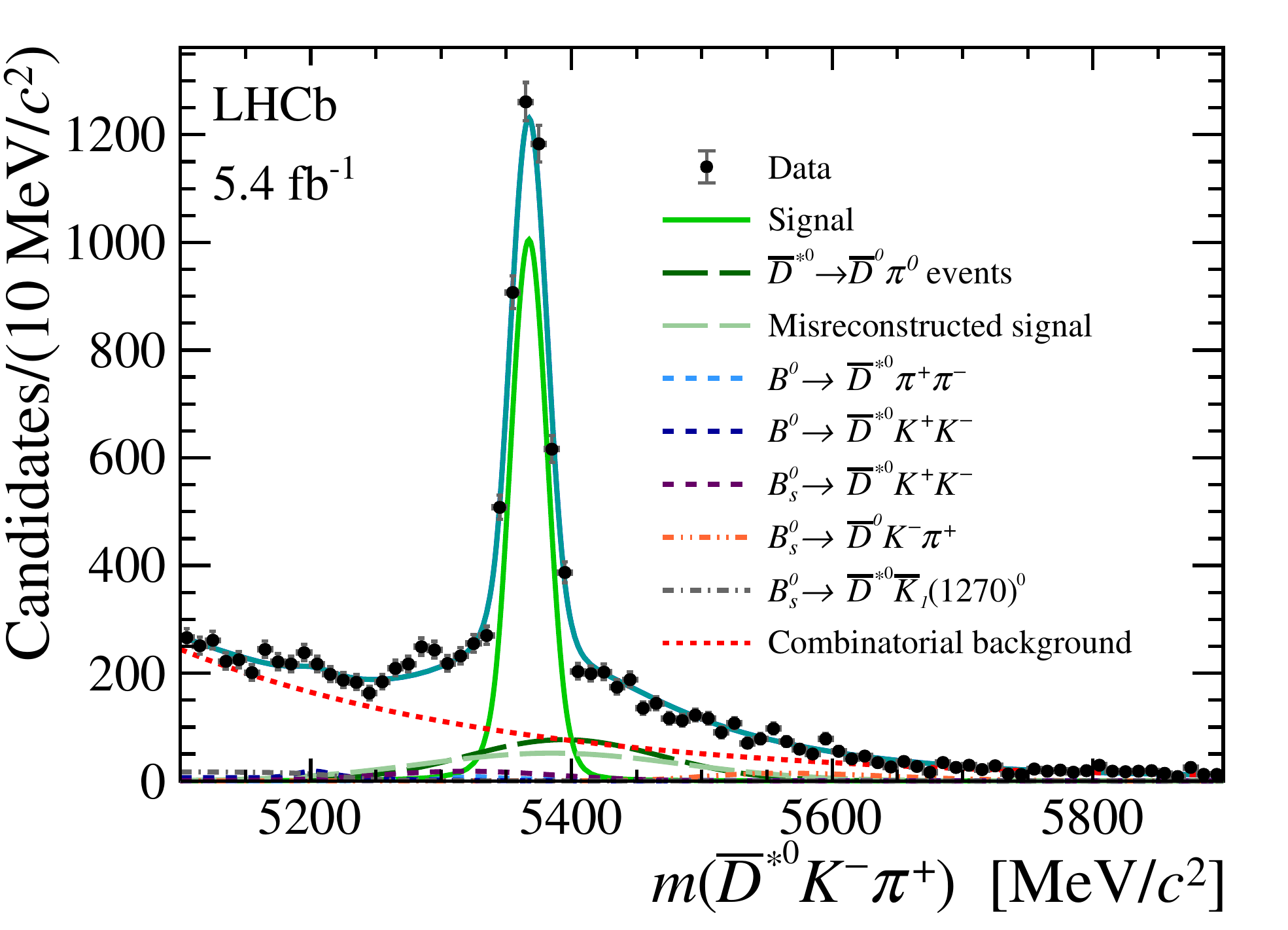}
  \includegraphics[width=0.48\linewidth]{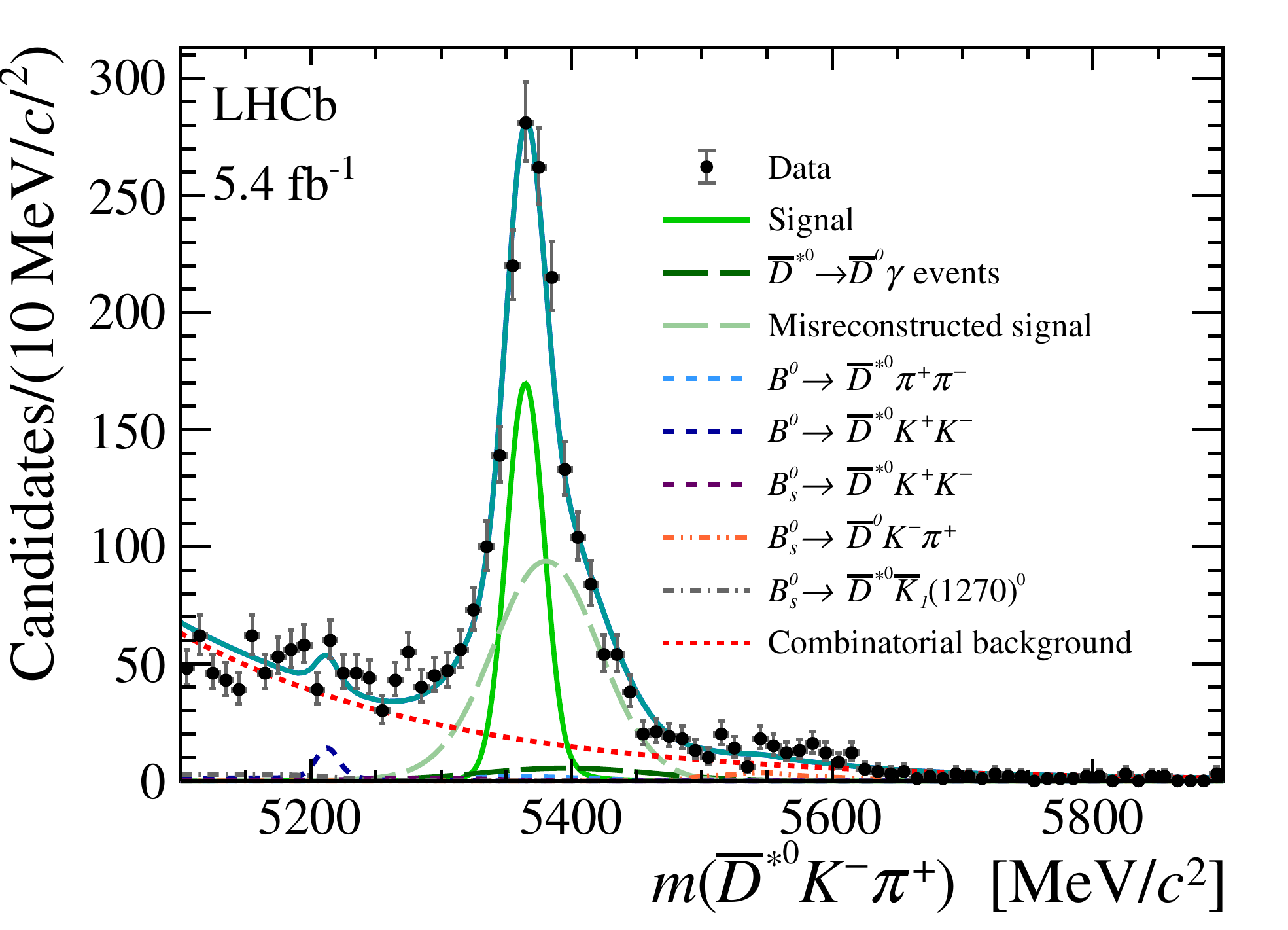}
  \includegraphics[width=0.48\linewidth]{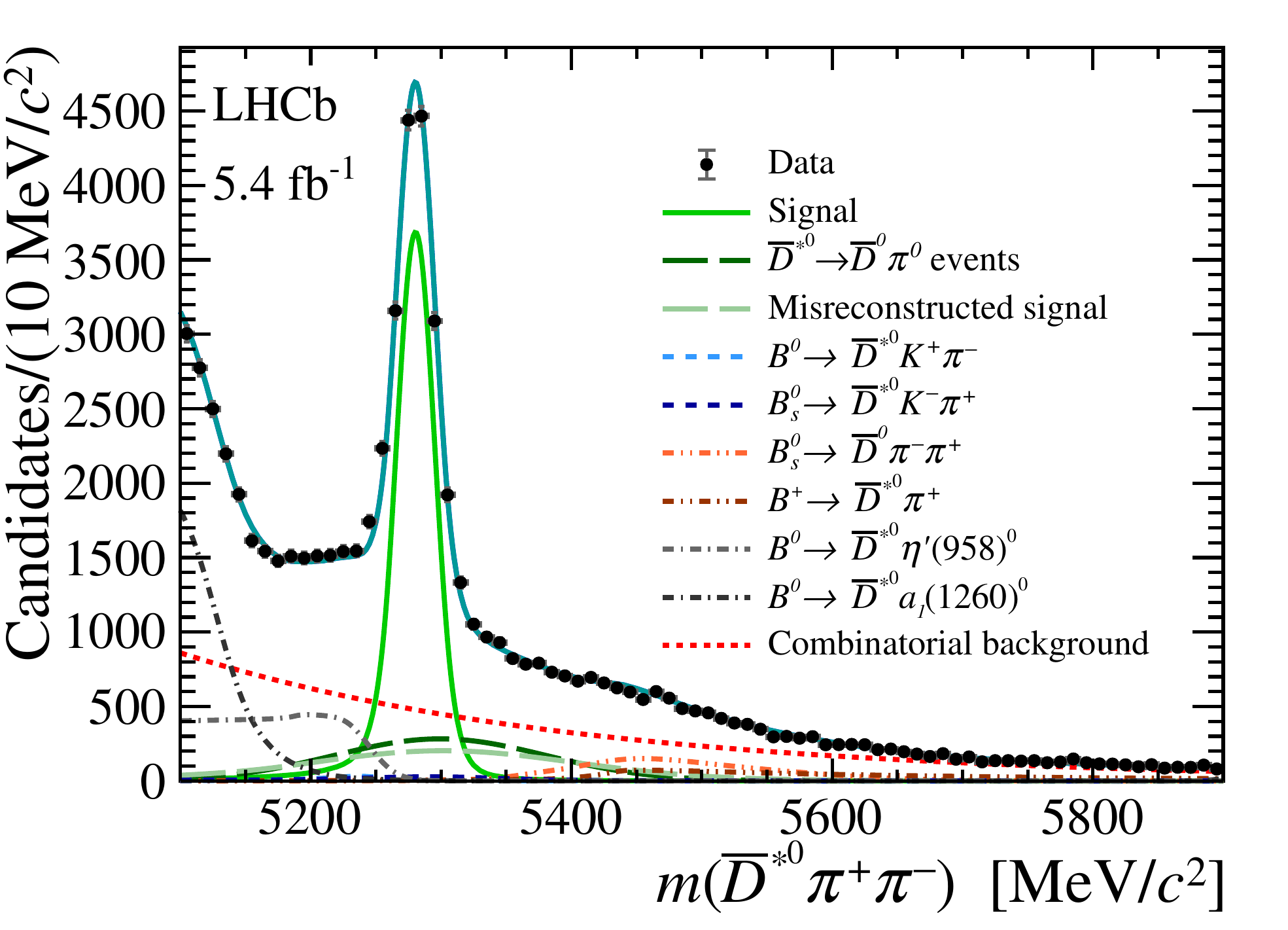}
  \includegraphics[width=0.48\linewidth]{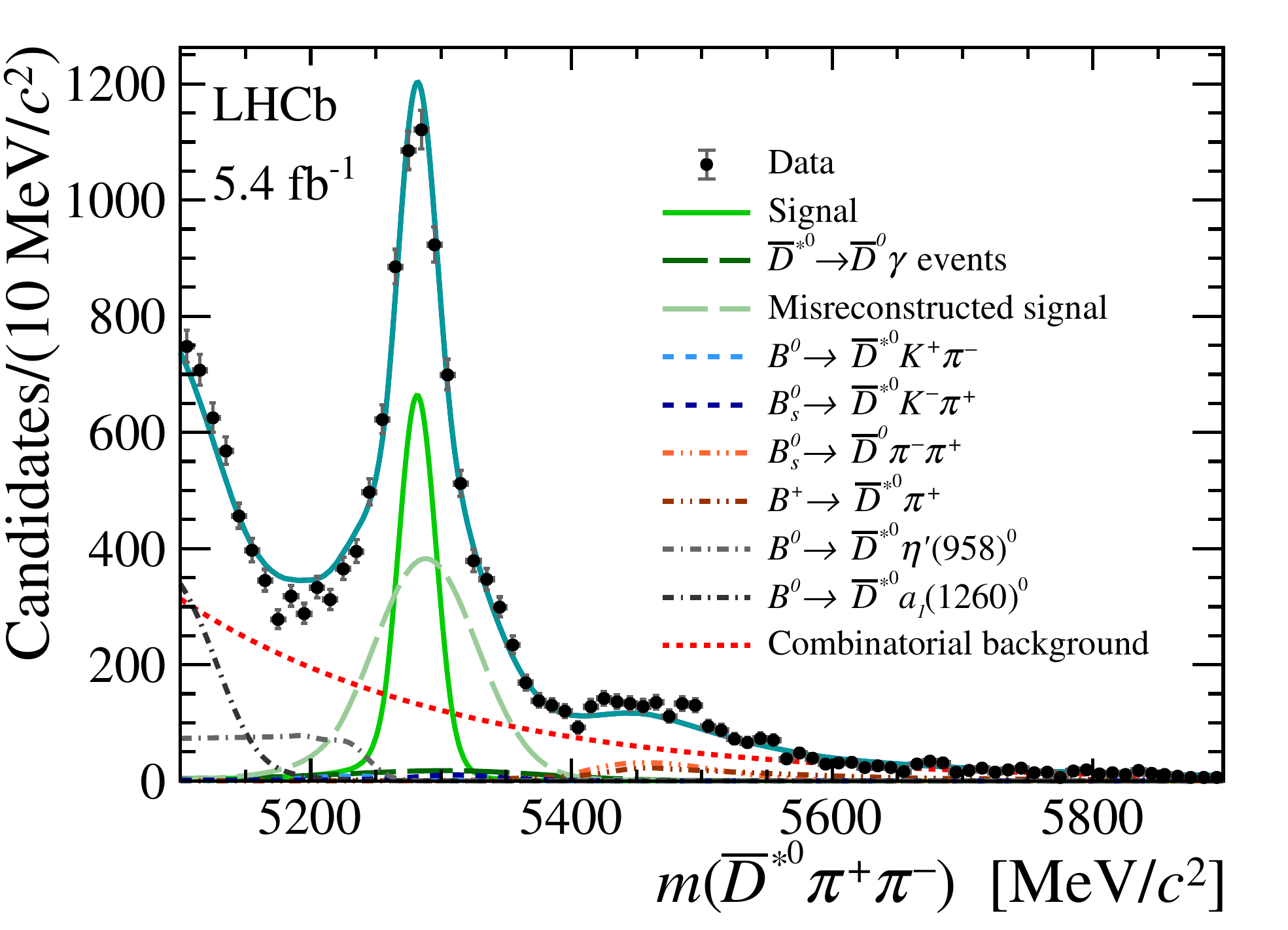}
  \caption{
    Invariant mass distributions of (top)~$B^0\to\Dstarzb \Kp\pim$, (middle)~$\Bs\to\Dstarzb \Km\pip$, and (bottom)~$B^0\to\Dstarzb \pip\pim$ with (left)~$\Dstarzb\to\Dzb\gamma$ and (right)~$\Dstarzb \to \Dzb\piz$ candidates, with linear $y$-axis.
    The total fit result is shown overlaid as a solid blue line, with individual components illustrated as indicated in the legends.
  }
  \label{fig:globalfit_lin}
\end{figure}

\begin{figure}[!tb]
  \centering
  \includegraphics[width=0.48\linewidth]{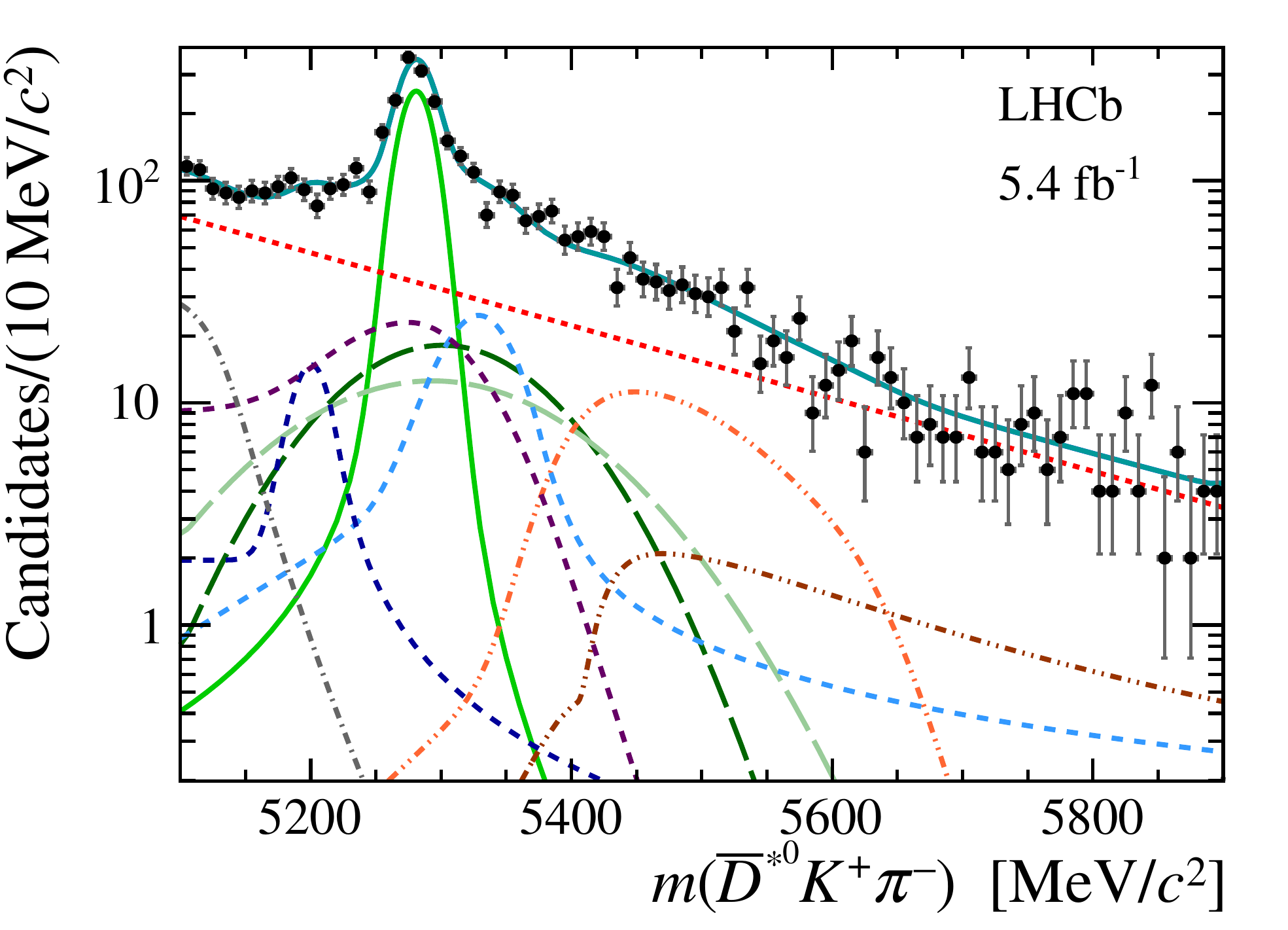}
  \includegraphics[width=0.48\linewidth]{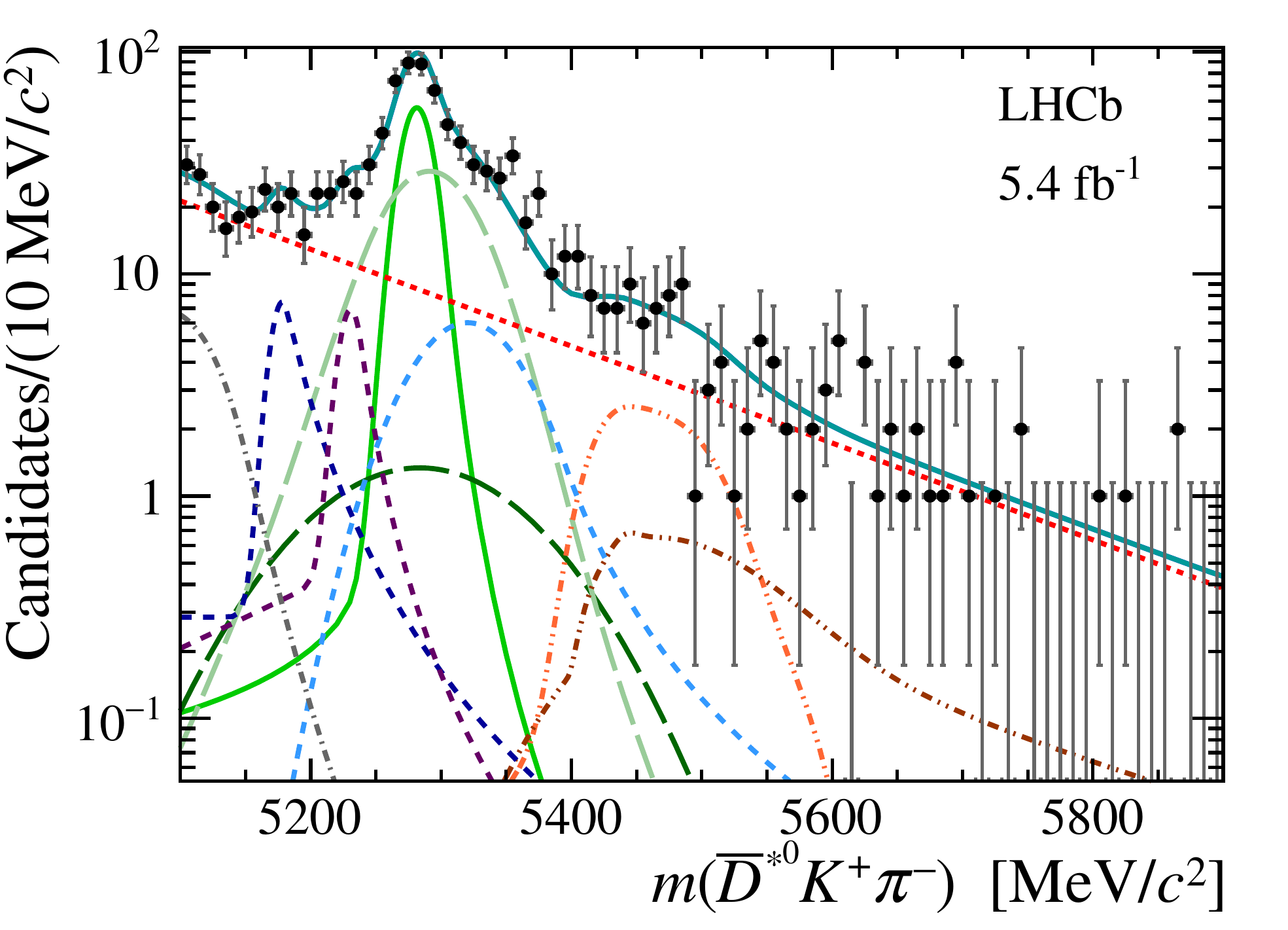}
  \includegraphics[width=0.48\linewidth]{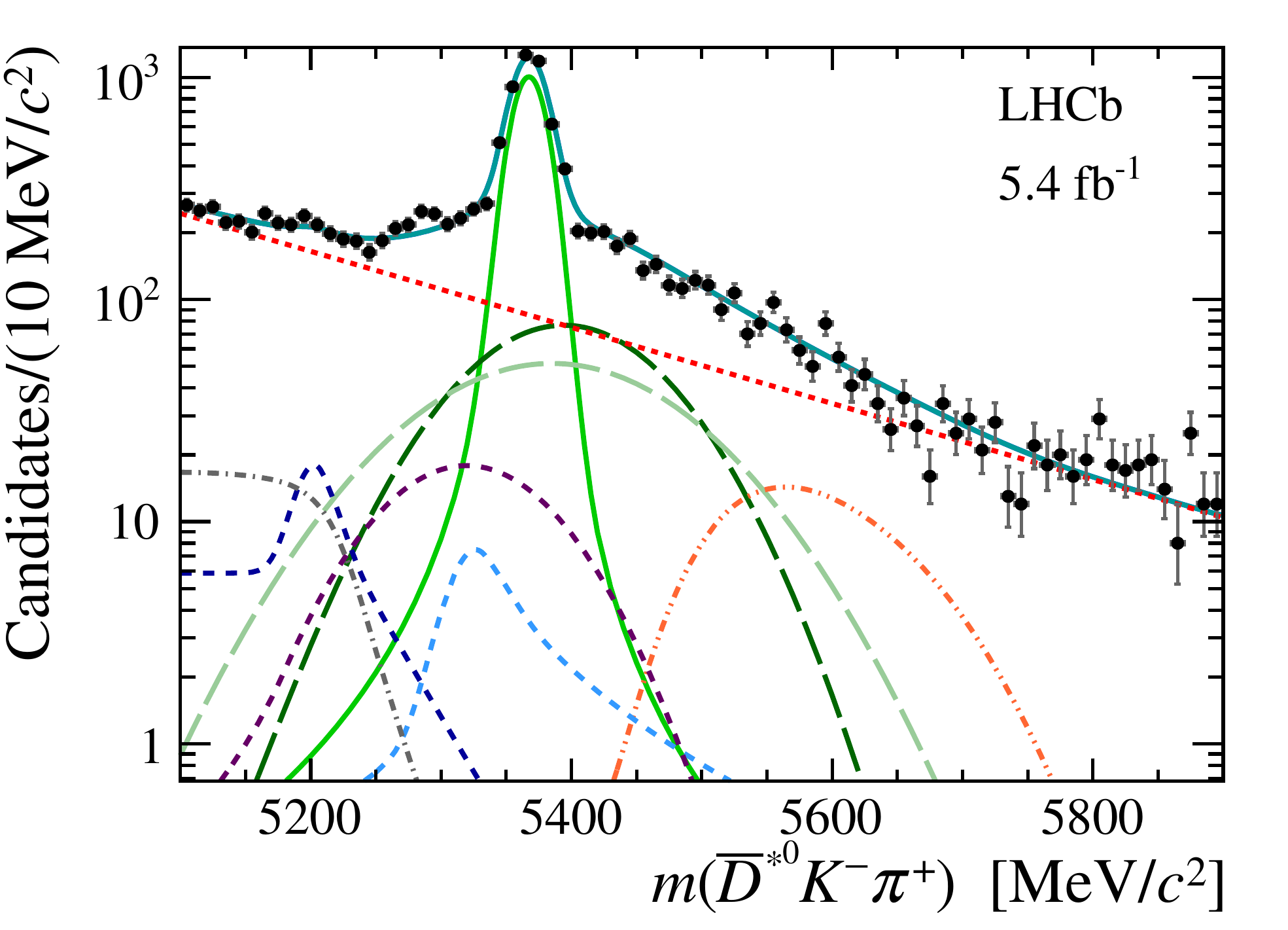}
  \includegraphics[width=0.48\linewidth]{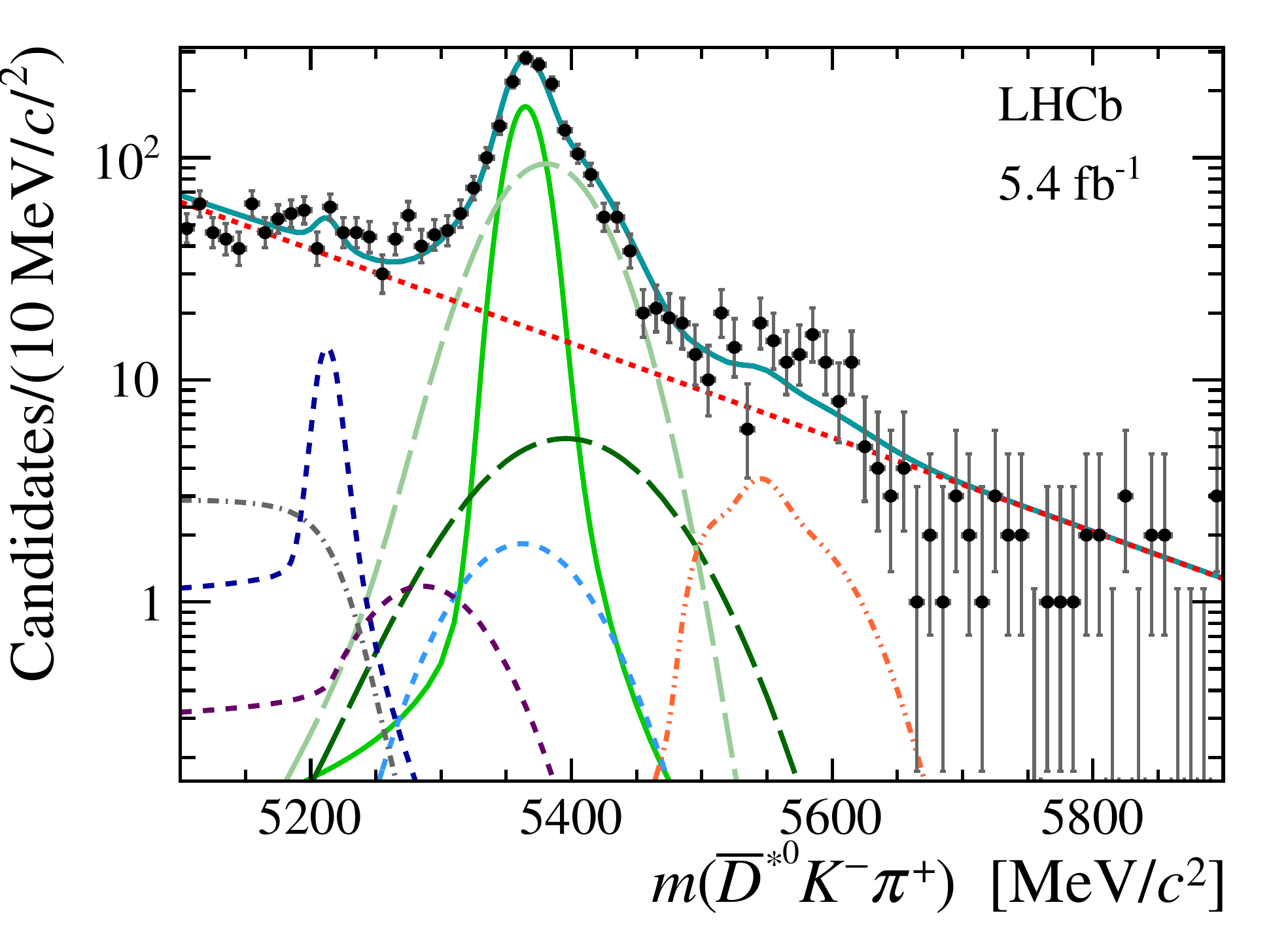}
  \includegraphics[width=0.48\linewidth]{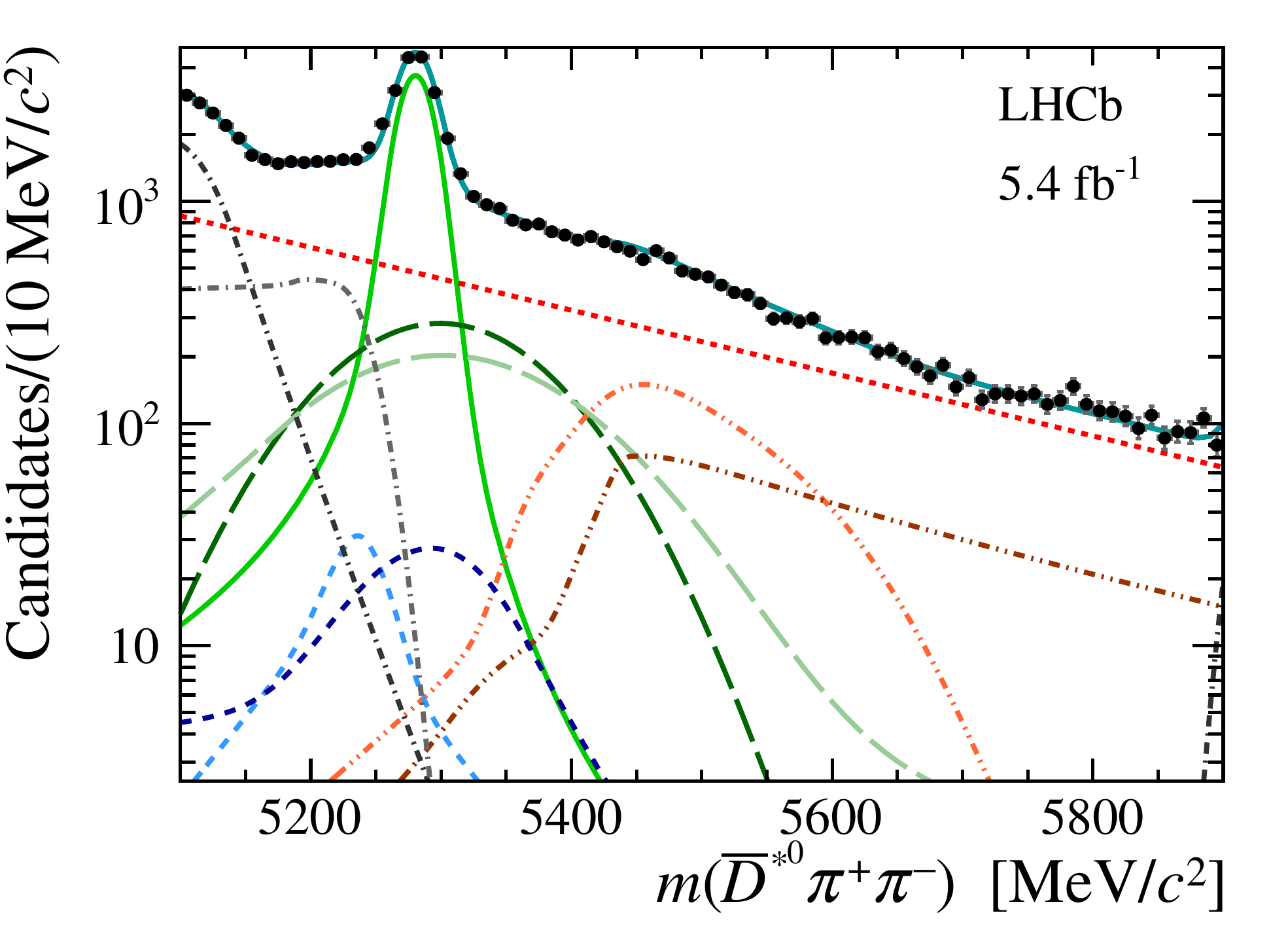}
  \includegraphics[width=0.48\linewidth]{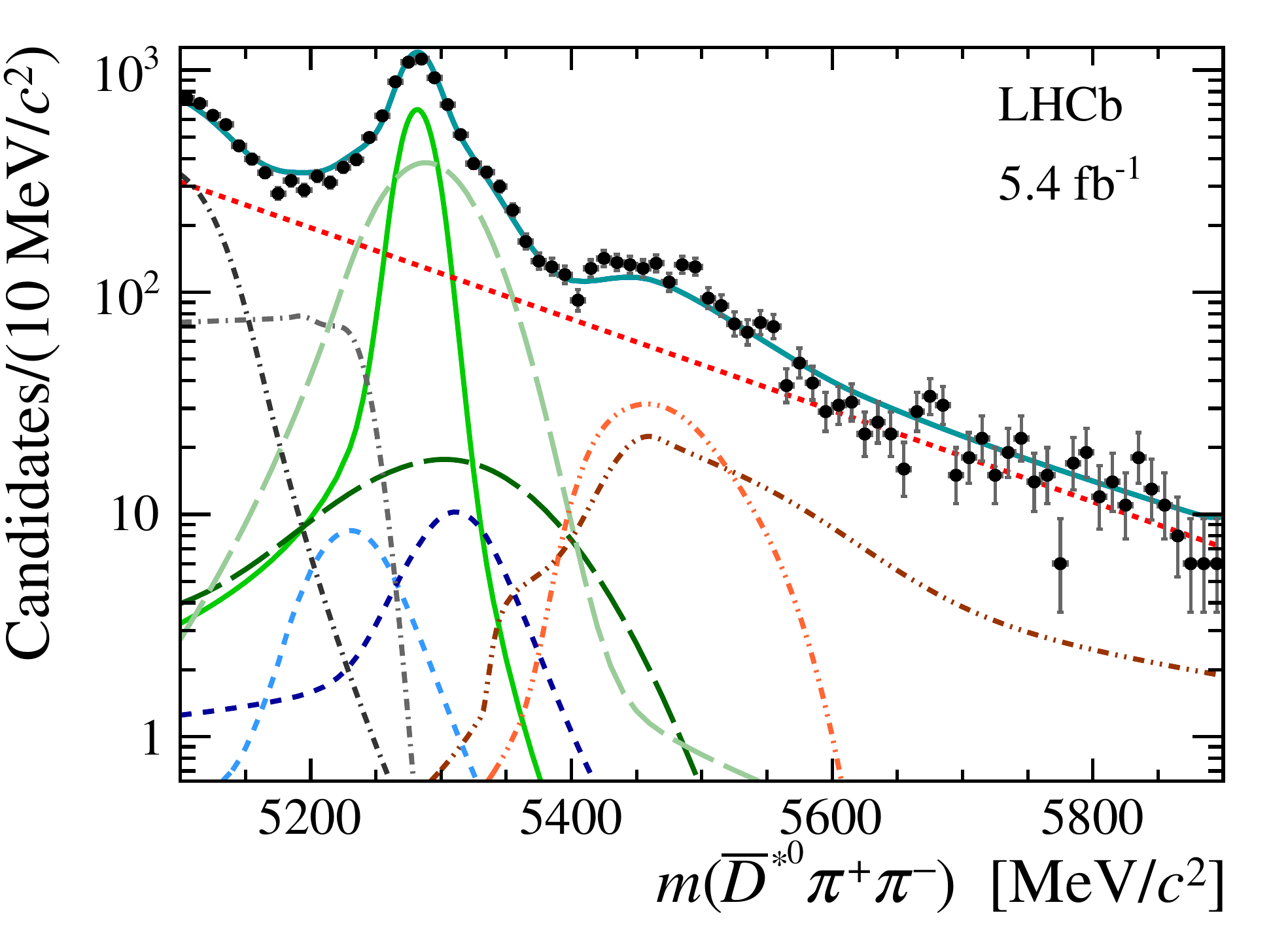}
  \caption{
    Invariant mass distributions of (top)~$B^0\to\Dstarzb \Kp\pim$, (middle)~$\Bs\to\Dstarzb \Km\pip$, and (bottom)~$B^0\to\Dstarzb \pip\pim$ with (left)~$\Dstarzb\to\Dzb\gamma$ and (right)~$\Dstarzb \to \Dzb\piz$ candidates, with logarithmic $y$-axis scale.
    The total fit result is shown overlaid as a solid blue line, with individual components illustrated as indicated in the legends of \cref{fig:globalfit_lin}.
  }
\label{fig:globalfit}
\end{figure}

The distributions of the \BdorBs-candidate invariant mass, $m(\Dstarzb h^+h^{\prime -})$, for the samples passing the selection requirements in each of the six final states are shown in Figs.~\ref{fig:globalfit_lin} and~\ref{fig:globalfit}.
These samples consist of several different components.
In order to determine the signal yield reliably, it is necessary to understand the different sources of these components, and their distributions in the \BdorBs-candidate mass spectra.
In several cases, the background sources form peaks in invariant mass distributions when reconstructed under the appropriate hypothesis and hence could in principle be vetoed. 
Such vetoes however tend to sculpt the signal efficiency variation across the phase space in a way that can be difficult to model, and can also distort the distributions of the remaining background components.  
To avoid potential bias due to these effects, these components are not vetoed and instead are modelled explicitly.

\subsection{Modelling of signal and background components}
\label{sec:massfit-components}

\paragraph{Signal and misidentified background:}
Simulation is used to study the detector response for both signal decays and misidentified background components.
Such background arises from \bquark-hadron decays to $\Dstarzb h^+h^{\prime -}$ final states, which pass the selection requirements with misidentification of the $h^+$ or $h^{\prime -}$ particles.
Misidentified background includes cases of cross-feed, where a decay that is signal in one final state forms a background to another.
Non-negligible contributions to the $\Dstarzb \Kpm\pimp$ final states are also found from $\Bd \to \Dstarzb \Kp\Km$ and $\Bs \to \Dstarzb \Kp\Km$ decays.
Contributions from $\Lbbar \to \Dstarzb \antiproton \Kp$ and $\Lbbar \to \Dstarzb \antiproton \pip$ decays cannot be ruled out but are not included in the baseline fit model, and instead are considered as a source of systematic uncertainty.

If the final-state hadrons are correctly identified, then the \BdorBs-candidate mass distribution does not depend strongly on the phase-space distribution of the decay.
When, however, a final-state particle is misidentified the \BdorBs-candidate mass is shifted by an amount that depends on the momentum of that particle, inducing a dependence of the shape on the Dalitz plot of the decay.  
Therefore, it is necessary to weight the simulation samples to give closer representations of the true distributions.
As the correlations between phase-space distribution and \BdorBs-candidate mass are not strong, it is sufficient to use models that give a reasonable, but not necessarily exact, representation of the true Dalitz plots.
For the $\Bd\to\Dstarzb\pip\pim$ decay this is done with a model, based on an analysis of the amplitude structure~\cite{Belle:2004}, that contains contributions from $\Dstarzb\rho(770)^0$, $\Dstarzb f_2(1270)$, $D_1(2420)^-\pip$ and $D_2^*(2460)^-\pip$.
For the other decays, no previous information is available on their phase-space distributions, but reasonable assumptions can be made based on data for the counterpart $\BdorBs\to\Dzb h^+h^{\prime -}$ decays~\cite{LHCb-PAPER-2012-018,LHCb-PAPER-2014-036,LHCb-PAPER-2015-017,LHCb-PAPER-2018-014,LHCb-PAPER-2018-015} and knowledge of the resonances that are likely to be present~\cite{PDG2020}.
The model for $\Bd\to\Dstarzb\Kp\pim$ decays includes contributions from $\Dstarzb K^*(892)^0$, $\Dstarzb K_0^*(1430)^0$, $D_1(2420)^-\Kp$ and $D_2^*(2460)^-\Kp$ transitions;
that for $\Bs\to\Dstarzb\Km\pip$ decays comprises $\Dstarzb \Kbar{}^*(892)^0$, $\Dstarzb \Kbar{}_0^*(1430)^0$, $D_{s1}(2536)^-\pip$ and $D_{s2}^*(2573)^-\pip$ components;
the model for $\Bd\to\Dstarzb\Kp\Km$ decays contains $\Dstarzb a_0(980)^0$, $\Dstarzb f_2^\prime(1525)$, $D_{s1}(2536)^-\Kp$ and $D_{s2}^*(2573)^-\Kp$ processes; while that for $\Bs\to\Dstarzb\Kp\Km$ decays has contributions from $\Dstarzb \phi(1020)$, $\Dstarzb f_2^\prime(1525)$, $D_{s1}(2536)^-\Kp$ and $D_{s2}^*(2573)^-\Kp$ decays.

It is possible for signal decays to be misreconstructed but still pass all selection requirements.
This can occur through misassociation of the soft neutral particle ($\gamma$ or $\piz$) produced in the \Dstarzb\ decay, in which case both \BdorBs-candidate mass and Dalitz-plot position are smeared to a larger extent than for correctly reconstructed signal.
This occurs more often for $\Dstarzb\to\Dzb\piz$ than for $\Dstarzb\to\Dzb\gamma$ decays due to the relatively high probability to replace one soft photon by another in the $\piz\to\gamma\gamma$ candidate and still pass the selection requirements.
The effect of misreconstruction on the $\Dstarzb-\Dzb$ candidate mass difference is shown in Fig.~\ref{fig:Dstarmass}.
It is also possible for a $\Dstarzb \to \Dzb\gamma$ decay to be reconstructed as a $\Dstarzb \to \Dzb\piz$ candidate, and vice versa.
These components are referred to as misreconstructed signal and wrong \Dstarzb\ decay, respectively, and are modelled separately from correctly reconstructed signal in order to avoid reliance on simulation of the misreconstruction rates.
In principle the components with charged particle misidentification could also be subdivided into \Dstarzb\ reconstruction categories, but this is found not to be necessary.

The correctly reconstructed signal, misreconstructed signal, wrong \Dstarzb\ decay, and misidentified background components are each modelled with the sum of two Crystal Ball functions~\cite{Skwarnicki:1986xj}, with a common peak position and width, and independent tails on opposite sides of the peak.
The shape parameters of these double Crystal Ball functions are determined from simulation and then fixed in the data fit, with the exception of an offset to the peak position and a scaling factor of the width that are common to all such shapes.

\paragraph{Partially reconstructed background:}
Partially reconstructed background originates from \bquark-hadron decays with a missing particle that is not included in the reconstruction of the signal candidate.
Numerous possible sources of partially reconstructed background are investigated with simulation~\cite{Cowan:2016tnm}, and it is found that the \BdorBs-candidate mass shape is similar for all cases with the same missing particle.
Thus $\Bz\to\Dstarzb K_1(1400)^0$, $\Bs\to\Dstarzb\Kbar{}_1(1400)^0$ and $\Bz\to\Dstarzb a_1(1260)^0$ decays are taken as proxies for the partially reconstructed background sources with a missing (charged or neutral) pion in the $\Dstarzb \Kp\pim$, $\Dstarzb \Km\pip$ and $\Dstarzb \pip\pim$ final states, respectively.
In addition, partially reconstructed $\Bz\to\Dstarzb \etapr(958), \etapr(958) \to \pip\pim\gamma$ decays, where the photon is missed, form a background to the $\Dstarzb \pip\pim$ final state.

The \BdorBs-candidate mass distribution for partially reconstructed background has a kinematic limit at $m_{X_b} - m_{\rm miss}$, where $m_{X_b}$ and $m_{\rm miss}$ are the mass of the decaying \bquark\ hadron and the missing particle, respectively.
The distribution extends to lower invariant mass values according to the momentum carried by the missing particle, resulting in a long tail on the low mass side of the distribution.
The underlying shape, with a kinematic limit and a long tail, is modelled by the ARGUS function~\cite{Albrecht:1990am}.
The distribution observed in data is smeared by the experimental resolution, so the kinematic limit does not appear as a hard edge.  
The resolution is expected to be similar to that for the signal decay, but since it is only significant for the high mass tail of the distribution, it can be described with a single, rather than a double, Crystal Ball function.

Partially reconstructed background components are therefore modelled by the convolution of an ARGUS function with a Crystal Ball function.
The parameters are determined from simulation and are subsequently fixed in the fit to data.

\paragraph{Combinatorial background:}
Combinatorial background arises from random combinations of particles, not originating from the same \bquark-hadron decay, that may be either correctly identified or misidentified.
Its shape in the \BdorBs-candidate invariant mass distribution is modelled with a falling exponential function.

\paragraph{Partially combinatorial background:}
A final category of background occurs where the majority of the particles comprising the \BdorBs\ candidate originate from a \bquark-hadron decay, combined with a random particle from the rest of the $pp$ collision event.
This is referred to as partially combinatorial background.
Such a background from $\BdorBs\to\Dzb\Kpm\pimp$ and $\Bd\to\Dzb\pip\pim$ decays is found to contribute significantly when combined with random soft neutral particles to form a \Dstarzb\ candidate.
Similarly, $\Bp\to\Dstarzb\Kp$ ($\Dstarzb\pip$) forms a background in the $\Dstarzb\Kp\pim$ ($\Dstarzb\pip\pim$) final state when combined with a random soft charged pion.

The shapes of these background components are studied with simulation and are found to be well-described by the sum of the Johnson function~\cite{Johnson} and a double Crystal Ball function.
Here, in contrast to the case for the signal and misidentified background components, the peak positions and width of the two Crystal Ball functions are allowed to differ.  
The parameters of these functions are obtained by fitting simulation samples and are then fixed in the fit to data, although the same peak position offset and width scaling factor are applied as for the signal and misidentified background components.
An exception is made for the partially combinatorial $\Bz\to\Dzb\pip\pim$ background to the $\Dstarzb\pip\pim$ final state that, due to its relatively large yield, requires particularly accurate modelling.
The \BdorBs-candidate mass distribution depends on the momentum spectrum of the soft neutral particles that are included in the candidate, which may differ between data and simulation.
Therefore, the shape of this component is obtained by selecting $\Bz\to\Dzb\pip\pim$ decays in data, based on the invariant mass of the candidate excluding the soft neutral particle.
The distribution obtained is fitted in the same way as the simulation.
A comparison of the shapes of this component in data and simulation shows reasonable agreement, validating the use of shapes from simulation for all other partially combinatorial components.

Another potential source of partially combinatorial background arises when a \mbox{$\Bm \to \theDstarm h^+h^{\prime -}$} decay is misreconstructed with the charged pion from the $\theDstarm\to\Dzb\pim$ decay not being included and a random $\gamma$ or $\piz$ candidate instead being used to form a $\Dstarzb$ candidate.
These decays are, however, suppressed by small CKM matrix elements, with the highest branching fraction of ${\cal O}(10^{-6})$ expected for $\Bm \to \theDstarm \Km\pip$~\cite{LHCb-PAPER-2015-054}, and are thus negligible.
Potential sources of background from decays to the same set of final state particles without any intermediate charm meson, or with two charm mesons, are also rendered negligible by the selection criteria.

\subsection{Simultaneous fit configuration}
\label{sec:massfit-config}

As discussed above, the six selected final states --- $\Bd \to \Dstarzb\Kp\pim$, $\Bs \to \Dstarzb\Km\pip$ and $\Bd \to \Dstarzb\pip\pim$, each with both $\Dstarzb \to \Dzb\gamma$ and $\Dzb\piz$ decays --- contain a large number of background components in addition to correctly reconstructed and misreconstructed signal.
The signal yields are therefore determined with a simultaneous unbinned extended maximum-likelihood fit to the six \BdorBs-candidate mass distributions, while exploiting various relations between the yields of the different components in order to ensure fit stability and accuracy.

The ratio between misreconstructed and correctly reconstructed signal yields is expected to depend only on the $\Dstarzb$ decay mode.
The yields of misreconstructed signal components are thus given by the product of the correctly reconstructed signal yield and a factor that is shared across \BdorBs\ decays but differs between $\Dstarzb\to\Dzb\gamma$ and $\Dstarzb\to\Dzb\piz$ decays.
In a similar way, common factors are used to relate the yields of signal components with the wrong $\Dstarzb$ decay and correctly reconstructed signal yields.
These factors are allowed to vary freely in the fit to data.

The yields of cross-feed background components, which are correctly reconstructed in one final state, but have charged hadron misidentification in another, are related to the correctly reconstructed signal yields by relative efficiencies, which are the same for both \Dstarzb\ decays.
The yields of background due to misidentified $\Bz\to\Dstarzb\Kp\Km$ decays in different final states are similarly related to each other, with their overall yield allowed to vary freely, and likewise for misidentified $\Bs\to\Dstarzb\Kp\Km$ decays.
The expected central values of the relative yields of the misidentified background components are determined from simulation and are constrained within their uncertainties in the fit to data.

The same partially reconstructed background sources are expected to be seen in final states that differ by only \Dstarzb\ decay, and hence the yields of these background components are fixed relative to each other.
These factors are allowed to vary freely in the fit to data.

The relative yields of partially combinatorial $\Bz\to\Dzb h^+h^{\prime -}$ background components between the different \Dstarzb\ final states depend only on the probability to form a $\Dstarzb$ candidate by combining with a random $\gamma$ or $\piz$ candidate.
Thus, this ratio of yields is expected to be the same for each $h^+h'^-$ combination.
Similarly, the yields of background from partially combinatorial $B^+\to\Dstarzb h^+$ decays depend on the probability to make a combination with a random charged pion, which is expected to be independent of the $\Dstarzb$ decay mode.
In addition, the relative yields of partially combinatorial background from $\Bu\to\Dstarzb\Kp$ and $\Bu\to\Dstarzb\pip$ decays can be constrained from their relative branching fractions, taken from Refs.~\cite{LHCb-PAPER-2017-021,Aubert:2004hu,LHCb-PAPER-2020-036}, and reconstruction probabilities.
To constrain further the partially combinatorial components, the yields of the $\Bu\to\Dstarzb\pip$ and $\Bz\to\Dzb\pip\pim$ contributions to the $\Bz\to\Dstarzb\pip\pim,\Dstarzb\to\Dzb\gamma$ channel are related to each other using knowledge of both the relative branching fractions~\cite{LHCb-PAPER-2017-021,Ablikim:2014mww,LHCb-PAPER-2014-070,PDG2020} and reconstruction probabilities.
For all these constraints, the central values and uncertainties are calculated using relative efficiencies determined from simulation. 
The relative yields are then constrained within these uncertainties in the fit to data.

In total, the simultaneous fit has 49 degrees of freedom, 12 of which are not completely floated but have Gaussian constraints applied.
The 37 freely varying parameters are: the 6 signal yields, 4 ratios of misreconstructed signal and wrong $\Dstarzb$ decays relative to correctly reconstructed signal, 6 combinatorial background yields, 6 combinatorial background slope parameters, 4 yields of misidentified background from $\BdorBs\to\Dstarzb\Kp\Km$ decays to the $\Dstarzb\Kp\pim$ final states, 3 yields of partially combinatorial $\BdorBs \to \Dzb h^+h'^-$ background components, 4 ratios related to the yields of partially reconstructed decays and 2 shift and 2 scale parameters that quantify differences in the signal shape between data and simulation.
The 12 parameters with external Gaussian constraints are composed of 6 that constrain the misidentified background yields, and 6 that constrain the partially combinatorial background yields.

\subsection{Fit results}
\label{sec:massfit-results}

Projections of the fit result are superimposed on the \BdorBs-candidate mass distributions shown in Figs.~\ref{fig:globalfit_lin} and~\ref{fig:globalfit}.
The fitted signal yields are given in \cref{tab:FitResults}, with their statistical correlations in \cref{tab:fitcorrelations}.
Due to the nature of the simultaneous fit, with various constraints between different components, it is expected that there may be significant correlations between fit parameters.
Large correlations, up to almost $90\%$, are seen between yields with $\Dstarzb \to \Dzb\pi^0$ decays.
This feature is investigated with pseudoexperiments and found to arise mainly as a consequence of the yield of misreconstructed, relative to correctly reconstructed, signal being constrained to be the same in the three final states with $\Dstarzb \to \Dzb\pi^0$ decays.
These correlations are taken into account when calculating the ratios of branching fractions, as described in \cref{sec:results}.

\begin{table}[!tb]
  \caption{
    Yields obtained from the simultaneous fit for the correctly reconstructed signal component.
    Uncertainties are statistical only.
  }
  \centering
  \def\arraystretch{1.2}
  \begin{tabular}{lr@{$\,\pm\,$}l}
    \hline
    Component & \multicolumn{2}{c}{Yield} \\
    \hline
    $\Bz\to\Dstarzb\Kp\pim,\Dstarzb\to\Dzb\gamma$ & $946$ & $53$ \\
    $\Bz\to\Dstarzb\Kp\pim,\Dstarzb\to\Dzb\piz$ & $185$ & $17$ \\
    $\Bs\to\Dstarzb\Km\pip,\Dstarzb\to\Dzb\gamma$ & $3744$ & $77$ \\
    $\Bs\to\Dstarzb\Km\pip,\Dstarzb\to\Dzb\piz$ & $633$ & $ 46$ \\
    $\Bz\to\Dstarzb\pip\pim,\Dstarzb\to\Dzb\gamma$  & $15021$ & $218$ \\
    $\Bz\to\Dstarzb\pip\pim,\Dstarzb\to\Dzb\piz$ & $2591$ & $190$ \\
    \hline
  \end{tabular}
  \label{tab:FitResults}
\end{table}

\begin{table}[!tb]
  \caption{
    Correlations between the signal yields, as obtained from the simultaneous fit, accounting for statistical uncertainties only.
    The individual components are labelled with a shorthand, following the same ordering (left to right, and top to bottom) as in \cref{tab:FitResults}.
    The symbols $\Dbar{}^{*0}_{\gamma}$ and $\Dbar{}^{*0}_{\piz}$ refer to $\Dstarzb\to\Dzb\gamma$ and $\Dstarzb\to\Dzb\piz$ decays, respectively.
  }
  \centering
  \def\arraystretch{1.2}
  \begin{tabular}{c|cccccc}
    \hline
    & $\Dbar{}^{*0}_{\gamma}\Kp\pim$ & $\Dbar{}^{*0}_{\piz}\Kp\pim$ & $\Dbar{}^{*0}_{\gamma}\Km\pip$ & $\Dbar{}^{*0}_{\piz}\Km\pip$ & $\Dbar{}^{*0}_{\gamma}\pip\pim$ & $\Dbar{}^{*0}_{\piz}\pip\pim$ \\
    \hline
    $\Dbar{}^{*0}_{\gamma}\Kp\pim$         & \phantom{-}---      & $\phantom{-}0.02$ & $\phantom{-}0.22$ & $\phantom{-}0.02$ & $\phantom{-}0.09$ & $\phantom{-}0.06$ \\
    $\Dbar{}^{*0}_{\piz}\Kp\pim$           & $\phantom{-}0.02$ & \phantom{-}---      & $\phantom{-}0.06$ & $\phantom{-}0.65$ & $-0.03$ & $\phantom{-}0.68$ \\
    $\Dbar{}^{*0}_{\gamma}\Km\pip$         & $\phantom{-}0.22$ & $\phantom{-}0.06$ & \phantom{-}---      & $\phantom{-}0.10$ & $\phantom{-}0.11$ & $\phantom{-}0.09$ \\
    $\Dbar{}^{*0}_{\piz}\Km\pip$           & $\phantom{-}0.02$ & $\phantom{-}0.65$ & $\phantom{-}0.10$ & \phantom{-}---      & $-0.04$ & $\phantom{-}0.87$ \\
    $\Dbar{}^{*0}_{\gamma}\pip\pim$ & $\phantom{-}0.09$ & $-0.03$ & $\phantom{-}0.11$ & $-0.04$ & \phantom{-}---      & $\phantom{-}0.00$ \\
    $\Dbar{}^{*0}_{\piz}\pip\pim$   & $\phantom{-}0.06$ & $\phantom{-}0.68$ & $\phantom{-}0.09$ & $\phantom{-}0.87$ & $\phantom{-}0.00$ & \phantom{-}--- \\
    \hline
    \end{tabular}
  \label{tab:fitcorrelations}
\end{table}

The result of the fit is seen to agree well with the data over most of the \BdorBs-candidate mass range in all six final states.
To quantify the agreement, the \chisq\ for each final state is compared to the number of bins in the histograms used.
Due to the nature of the simultaneous fit, it is not possible to determine an appropriate number of degrees of freedom for each final state individually.
These $\chisq/N_{\rm bins}$ values, which account only for statistical uncertainties, are found to be in an acceptable range for the three final states with lowest yields, but larger values ($\chisq/N_{\rm bins} \approx 2$) are seen in the final states with higher numbers of selected candidates.
There is, however, no clear and consistent pattern of deviations between the data and the fit model between the two \Dstarzb\ decays in each final state.
Therefore, it is not possible to identify any further components that should be included in the fit, and the residual discrepancies are interpreted as arising from inaccuracies in the modelling, which are accounted for as a source of systematic uncertainty.

\section{Signal weighting and efficiency correction}
\label{sec:efficiency}

\begin{figure}[!tb]
  \centering
  \includegraphics[width=0.49\textwidth]{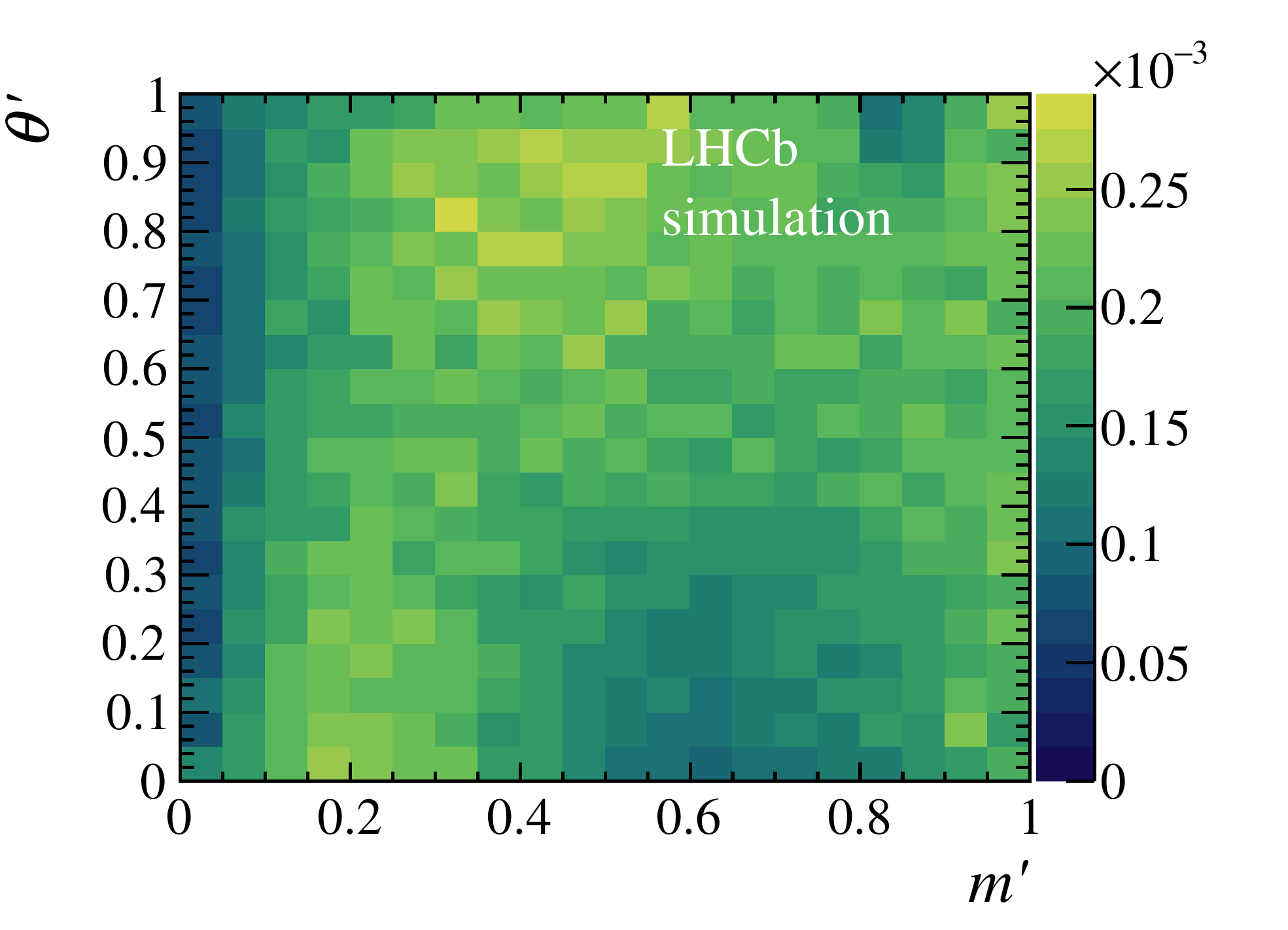}
  \includegraphics[width=0.49\textwidth]{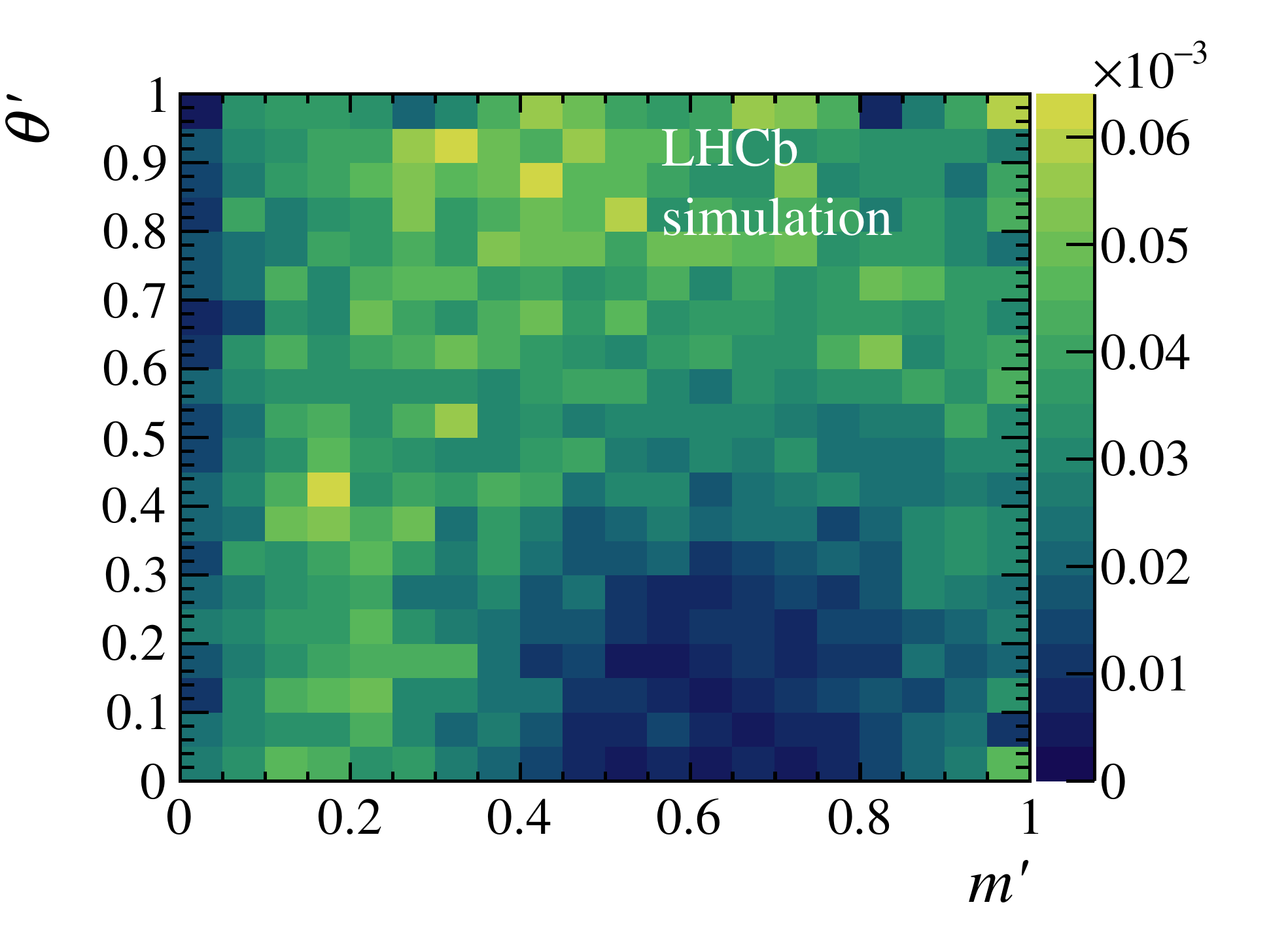}
  \includegraphics[width=0.49\textwidth]{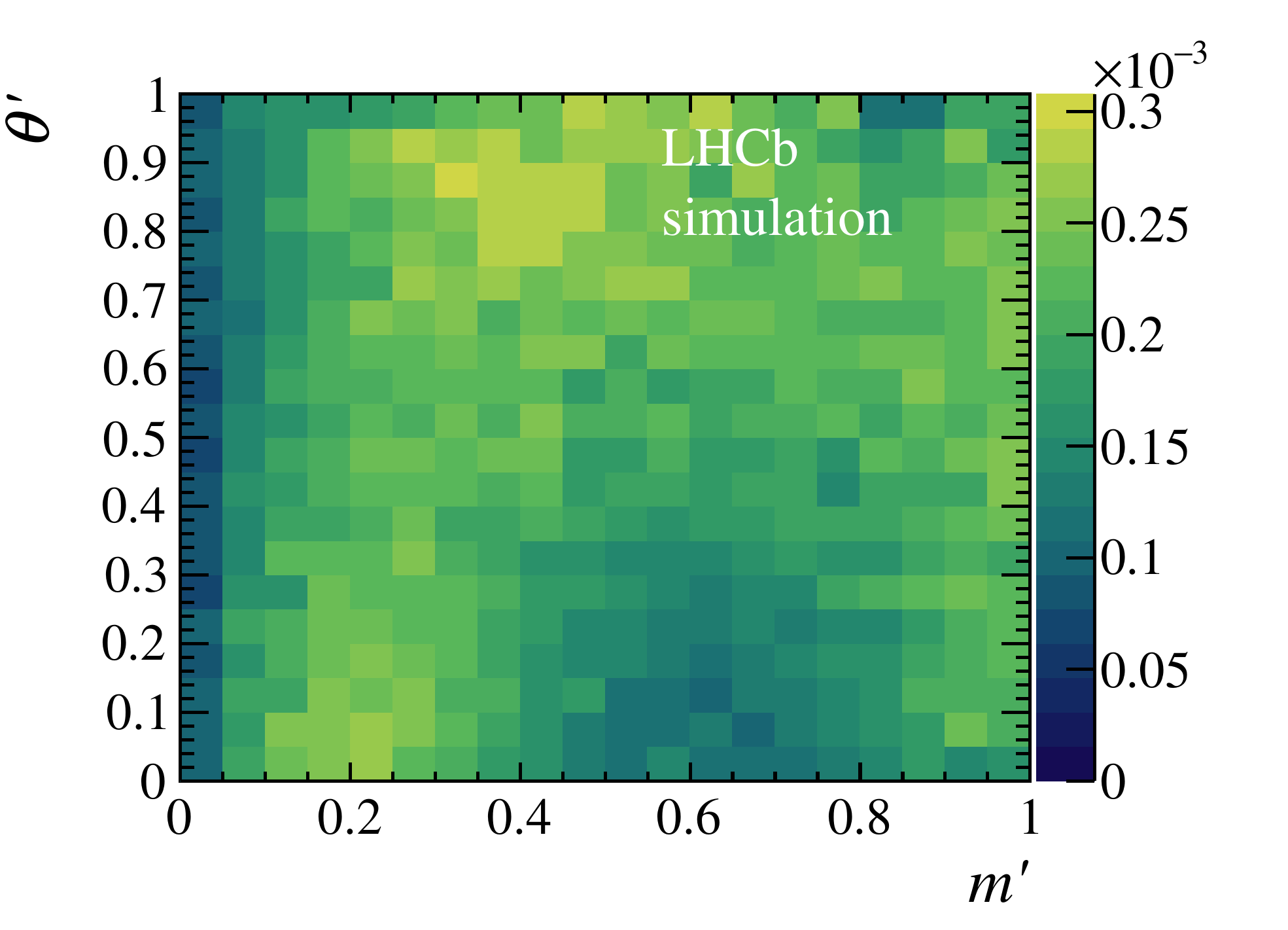}
  \includegraphics[width=0.49\textwidth]{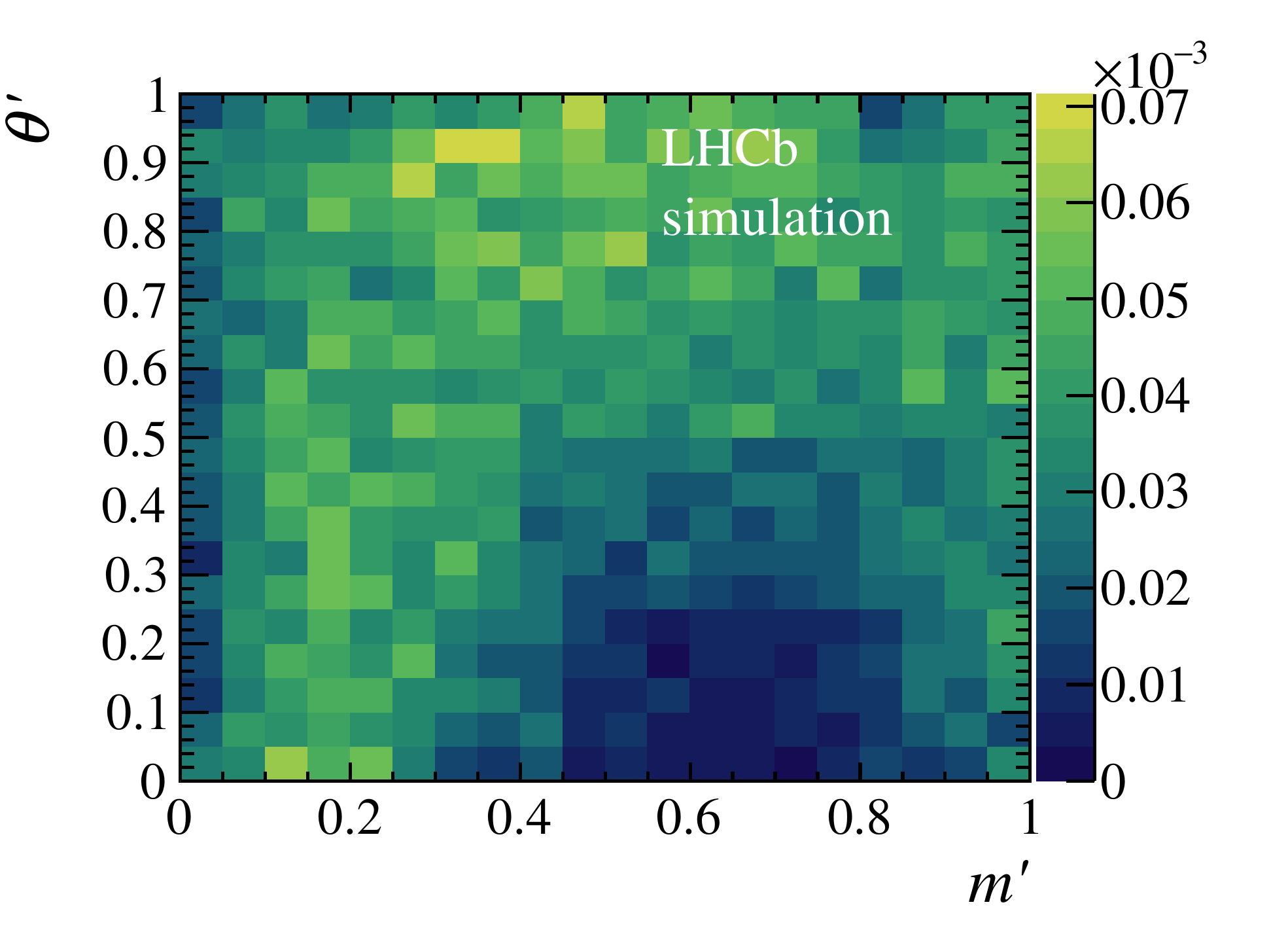}
  \includegraphics[width=0.49\textwidth]{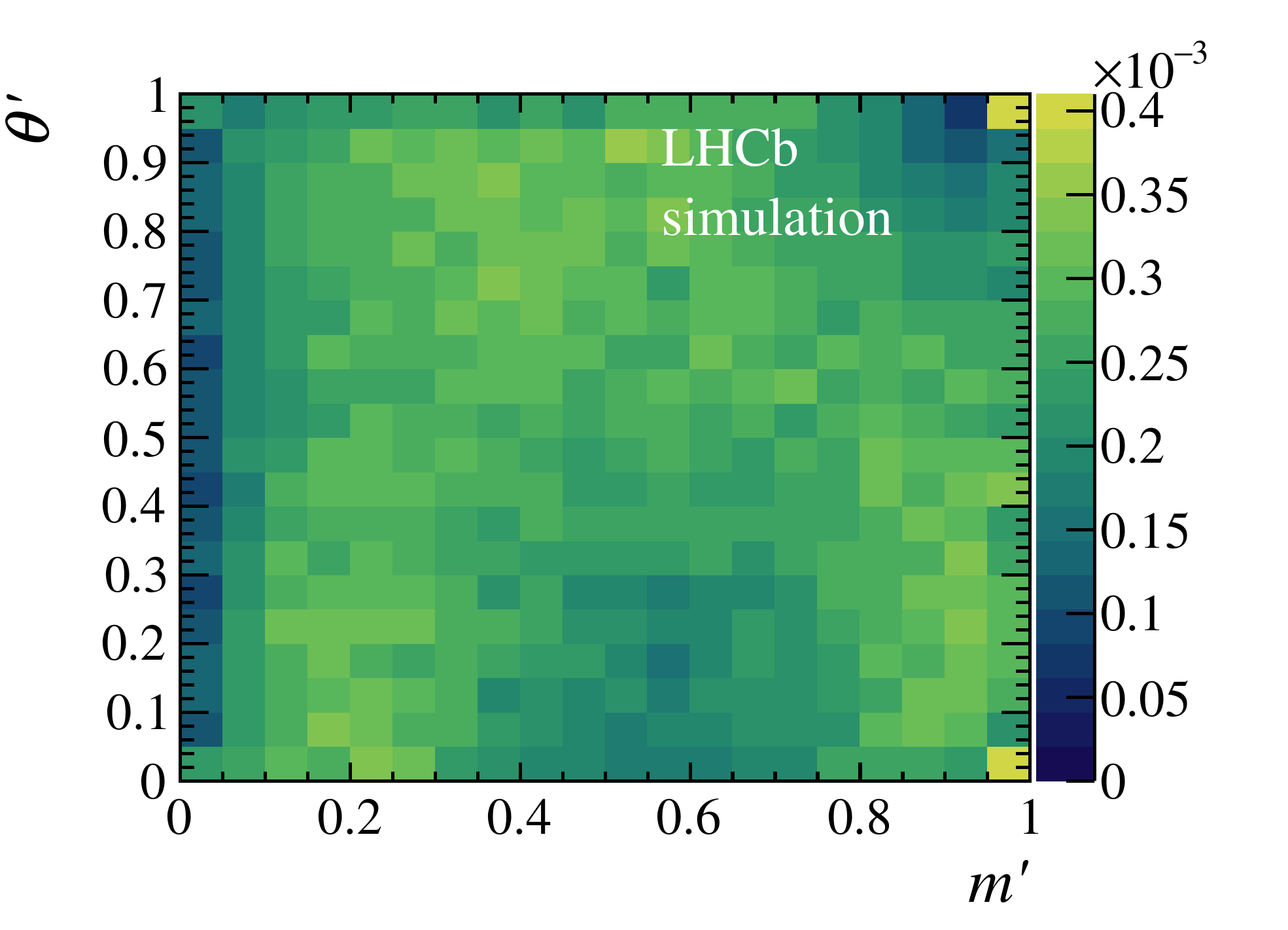}
  \includegraphics[width=0.49\textwidth]{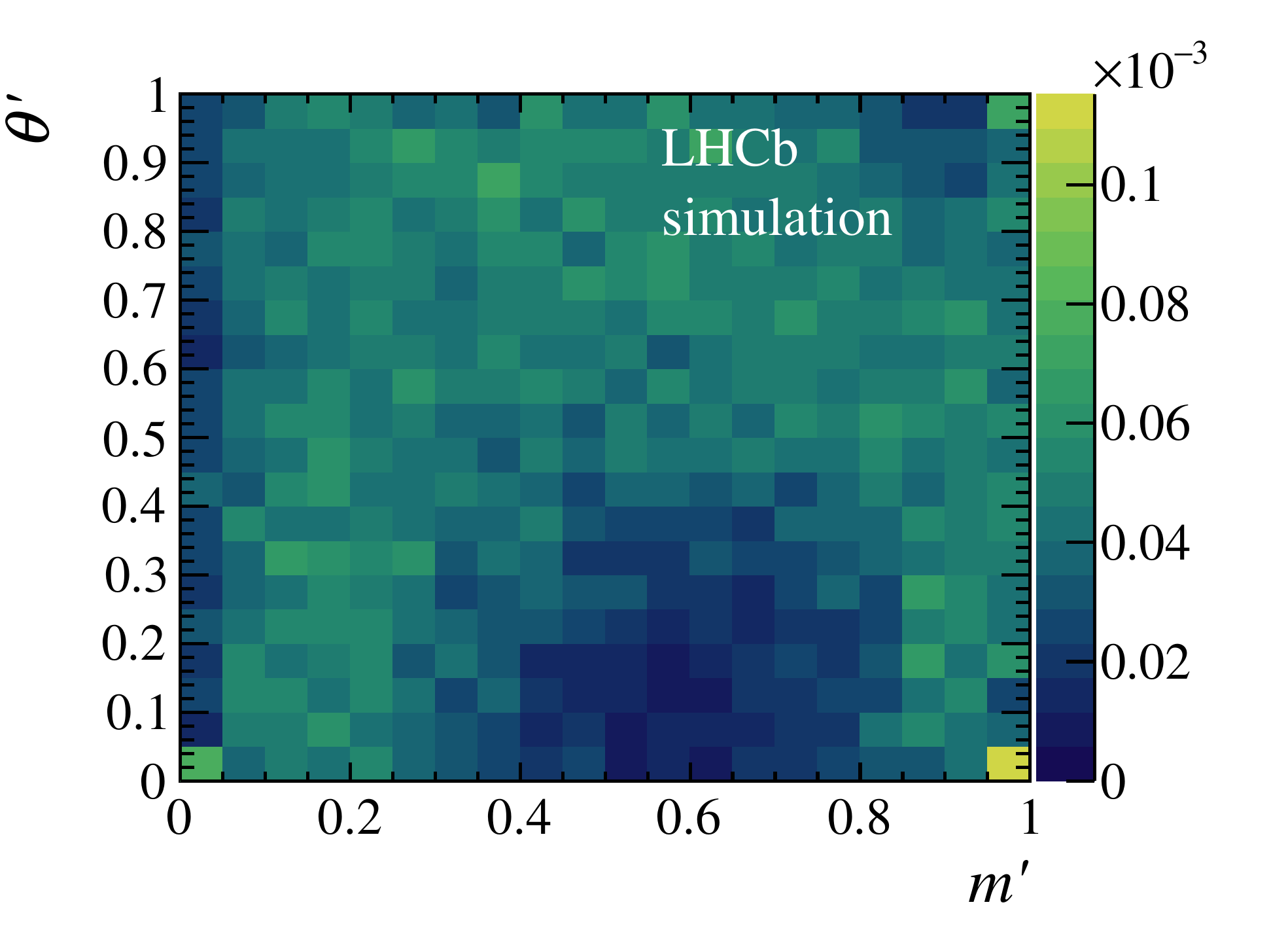}
  \caption{
    Total efficiency as a function of SDP coordinates for (top)~$\Bz\to\Dstarzb\Kp\pim$, (middle)~\mbox{$\Bs\to\Dstarzb\Km\pip$}, and (bottom)~$\Bz\to\Dstarzb\pip \pim$, with (left)~$\Dstarzb\to\Dzb\gamma$ and (right)~$\Dstarzb\to\Dzb\piz$ decays.
  }
  \label{fig:MCeffs}
\end{figure}

There are several effects contributing to the efficiency to detect a \BdorBs\ meson that is produced within the LHCb acceptance and decays to one of the final states of interest.
These are: geometrical acceptance, in which one or more of the final state particles passes outside the LHCb detector; trigger efficiency, both at hardware and software level, where the latter includes reconstruction effects; selection efficiency, of both prefiltering and final selection stages, which is the probability that a signal decay passes the requirements imposed in order to reduce background.
The efficiencies are determined from simulation in which, as noted in Sec.~\ref{sec:selection}, the values in simulation of variables used for charged hadron identification are drawn from data control samples.
As the simulation has been tuned to describe the kinematics of \BdorBs-meson production within the LHCb acceptance and the detector response, no further corrections are necessary.
The effects of acceptance, trigger efficiency, reconstruction, online selection and prefiltering are determined separately from those of offline selection in order to make efficient use of the computing resources available to produce simulation samples.
The total efficiency for each of the six signal modes is shown in \cref{fig:MCeffs}.

The simulation samples are generated flat across the square Dalitz plot (SDP) representation of the phase space.
The SDP is defined by the variables $m^\prime$ and $\theta^\prime$, both defined in the range $[0,1]$ and given by (see, \eg, Ref.~\cite{Back:2017zqt}): 
\begin{equation}
  m^\prime  =\frac{1}{\pi}\text{arccos}\left(2\frac{m_{12}-m_1-m_2}{m_B-m_1-m_2-m_3}-1\right) \, , \qquad
  \theta^\prime  = \frac{1}{\pi}\theta_{12} \, ,
\end{equation}
where $m_i$ is the  mass of the particle numbered $i$ (and equivalently for $m_B$), $m_{12}$ is the invariant mass of the 1-2 pair, and $\theta_{12}$ is the helicity angle of the 1-2 system, \ie\  the angle between the momenta of the particle numbered 1 and that numbered 3 in the rest frame of the 1-2 pair.
The values of $m_{12}$ and $\theta_{12}$ are obtained from a kinematic fit in which the $\BdorBs$, $\Dstarzb$ and $\Dzb$ masses are fixed to their known values~\cite{PDG2020}.
The definition is, however, dependent on the ordering of the particles; the specific ordering used to define the variables in \cref{fig:MCeffs} is given in \cref{tab:SDPconvention}.

\begin{table}[!tb]
  \caption{
    Ordering convention used to define the SDP representation of the phase space for each of the signal decays.
  }
  \centering
  \def\arraystretch{1.2}
  \begin{tabular}{lccc}
    \hline
    Channel & $1$ & $2$ & $3 $\\
    \hline
    $\Bz\to \Dstarzb \Kp \pim$ & $\Dstarzb$ & $\Kp$ & $\pim$ \\
    $\Bs\to \Dstarzb \Km \pip$ & $\Dstarzb$ & $\Km$ & $\pip$ \\
    $\Bz\to \Dstarzb \pip \pim$ & $\Dstarzb$ & $\pim$ & $\pip$ \\
    \hline
  \end{tabular}
  \label{tab:SDPconvention}
\end{table}

Since the signal decays involve various intermediate resonances, and the efficiency varies significantly across the phase space, it is necessary to use a weighting scheme to ensure that the simulation is appropriately matched to the data.
This has been done in many previous analyses (see, \eg, Ref.~\cite{LHCb-PAPER-2012-018}), but the standard \sPlot\ procedure~\cite{Pivk:2004ty} requires the use of weights that are obtained from a fit in which the only freely varying parameters are the yields of the different components.
Due to the nature of the simultaneous fit described in 
Sec.~\ref{sec:massfits}, in particular the application of constraints to certain background yields, that is not possible here.
Therefore, weights are instead obtained from a reformulation of the \sPlot\ approach~\cite{cows}, where no such requirement is necessary.
These weights can be used, for each category, to obtain a projection of the distribution of any variable which is independent of the \BdorBs-candidate mass.
This is, to a reasonable approximation, the case for the SDP and any other representation of the phase space of the signal decays.
Possible violations of this assumption of independence between the SDP and \BdorBs-candidate mass are considered as a source of systematic uncertainty in the results.

The signal weights and SDP efficiency maps are used to determine the ratio of branching fractions (illustrated here for generic decay modes $X$ and $Y$), as 
\begin{equation}
  \label{eq:effWeight}
  \frac{{\cal B}(X)}{{\cal B}(Y)} =
  \frac{\sum_i w_{X}(m_{B\,i})/\epsilon_X(m^{\prime}_i,\theta^{\prime}_i)}{\sum_j w_{Y}(m_{B\,j})/\epsilon_Y(m^{\prime}_j,\theta^{\prime}_j)} =
  \frac{N(X)/\left<\epsilon(X)\right>}{N(Y)/\left<\epsilon(Y)\right>} \,,
\end{equation}
where the indices $i$ and $j$ run over the $X$ and $Y$ candidates, respectively.
The functions $w_{X\,(Y)}$ and $\epsilon_{X\,(Y)}$ are the signal weight and efficiency function for $X\,(Y)$, respectively, and $m_{B\,i\,(j)}$, $m^{\prime}_{i\,(j)}$ and $\theta^{\prime}_{i\,(j)}$ are the \BdorBs-candidate mass and SDP variables for candidate $i\,(j)$.
Since the yield corresponds to the sum of the signal weights, \ie\ $N(X) = \sum_i w_{X}(m_{B\,i})$, the average weighted efficiency is given by
\begin{equation}
  \label{eq:avEff}
  \left<\epsilon(X)\right> = \frac{\sum_i w_{X}(m_{B\,i})}{\sum_i w_{X}(m_{B\,i})/\epsilon_X(m^{\prime}_i,\theta^{\prime}_i)} \, .
\end{equation}
For the ratio ${\cal B}(\Bs\to\Dstarzb\Km\pip)/{\cal B}(\Bz\to\Dstarzb\pip\pim)$ the right-hand side of Eq.~\eqref{eq:effWeight} is modified to include a factor of $\left(f_s/f_d\right)^{-1}$, where $f_s/f_d = 0.2539\pm0.0079$~\cite{LHCb-PAPER-2020-046} is the ratio of fragmentation fractions for $\Bs$ and $\Bz$ mesons in the LHCb acceptance.

The main goal of the analysis is the determination of the relative branching fractions of different \BdorBs\ decays to $\Dstarzb h^+h^{\prime -}$ final states.
However, Eq.~\eqref{eq:effWeight} can also be used to determine the ratio ${\cal B}(\Dstarzb\to\Dzb\gamma)/{\cal B}(\Dstarzb\to\Dzb\piz)$, which provides a useful cross-check of the procedures.
The results for this ratio are $0.53\pm0.06$, $0.59\pm0.04$ and $0.54\pm0.04$ in $\Bz\to\Dstarzb\Kp\pim$, $\Bs\to\Dstarzb\Km\pip$ and $\Bz\to\Dstarzb\pip\pim$ decays, respectively, where the uncertainties are statistical only.
These values are self-consistent and also agree with the most precise determinations of this ratio~\cite{Ablikim:2014mww,LHCb-PAPER-2017-021,LHCb-PAPER-2020-036}.

\begin{figure}[!htb]
  \centering
  \includegraphics[width=0.49\textwidth]{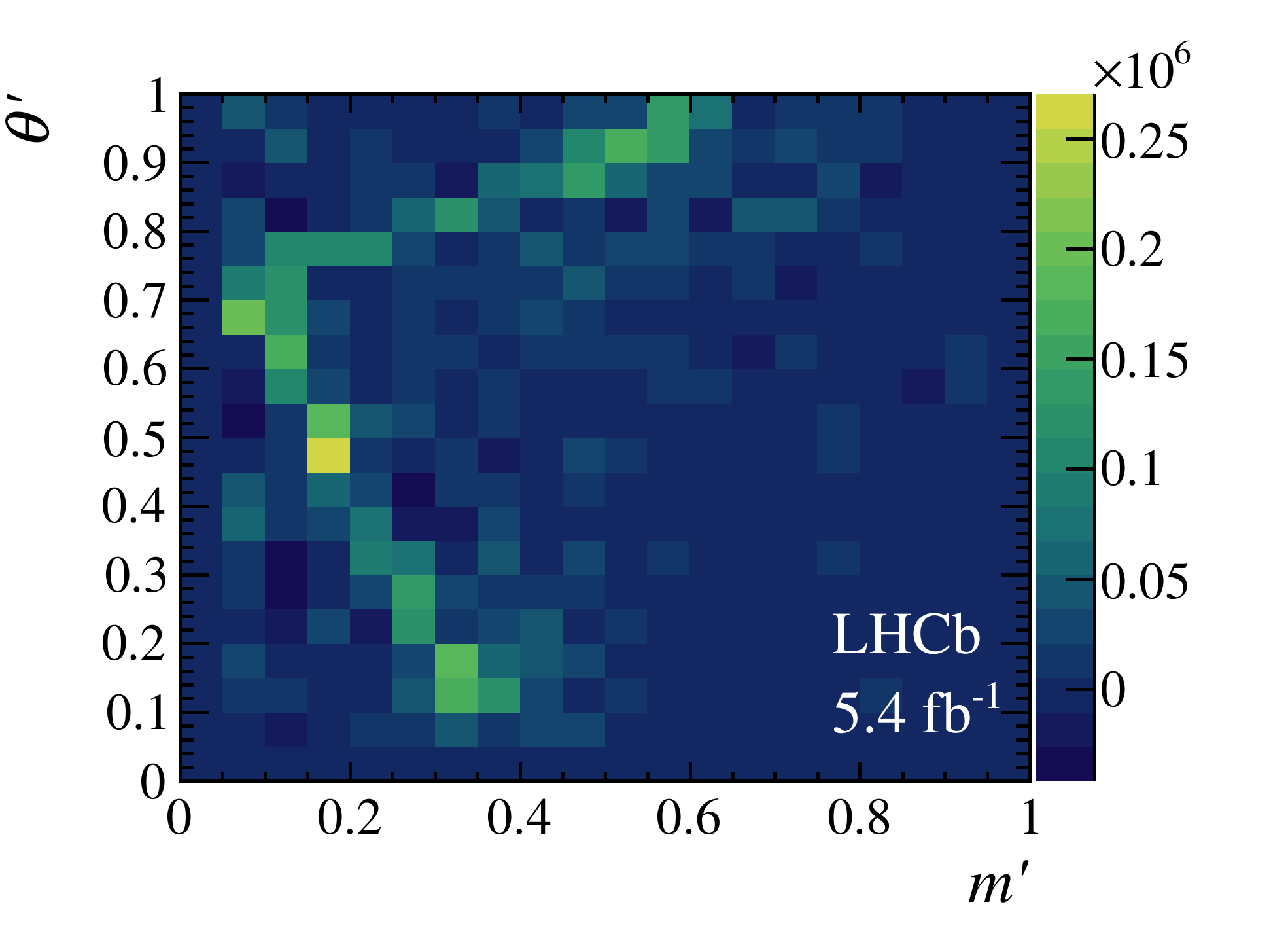}
  \includegraphics[width=0.49\textwidth]{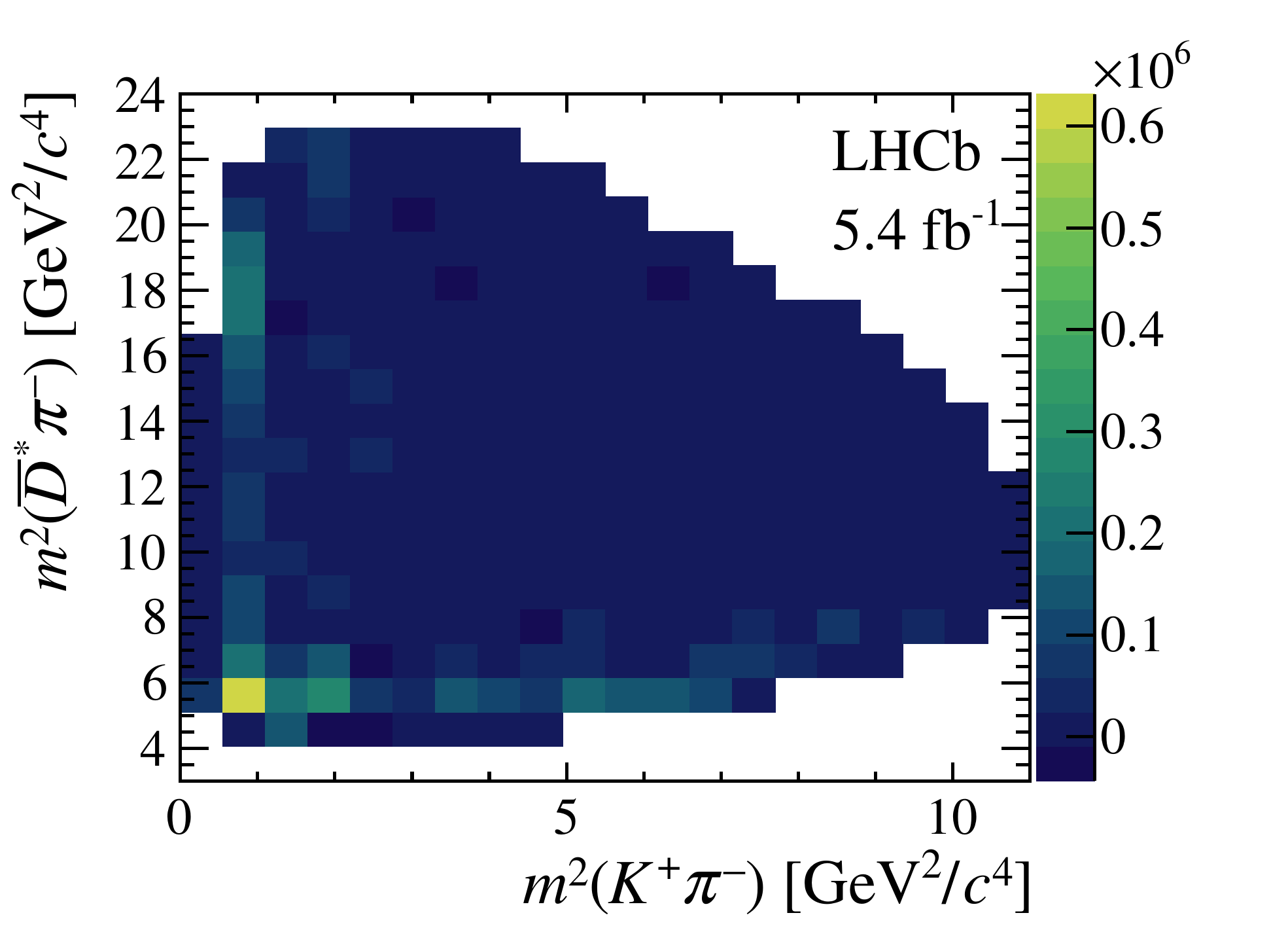}
  \includegraphics[width=0.49\textwidth]{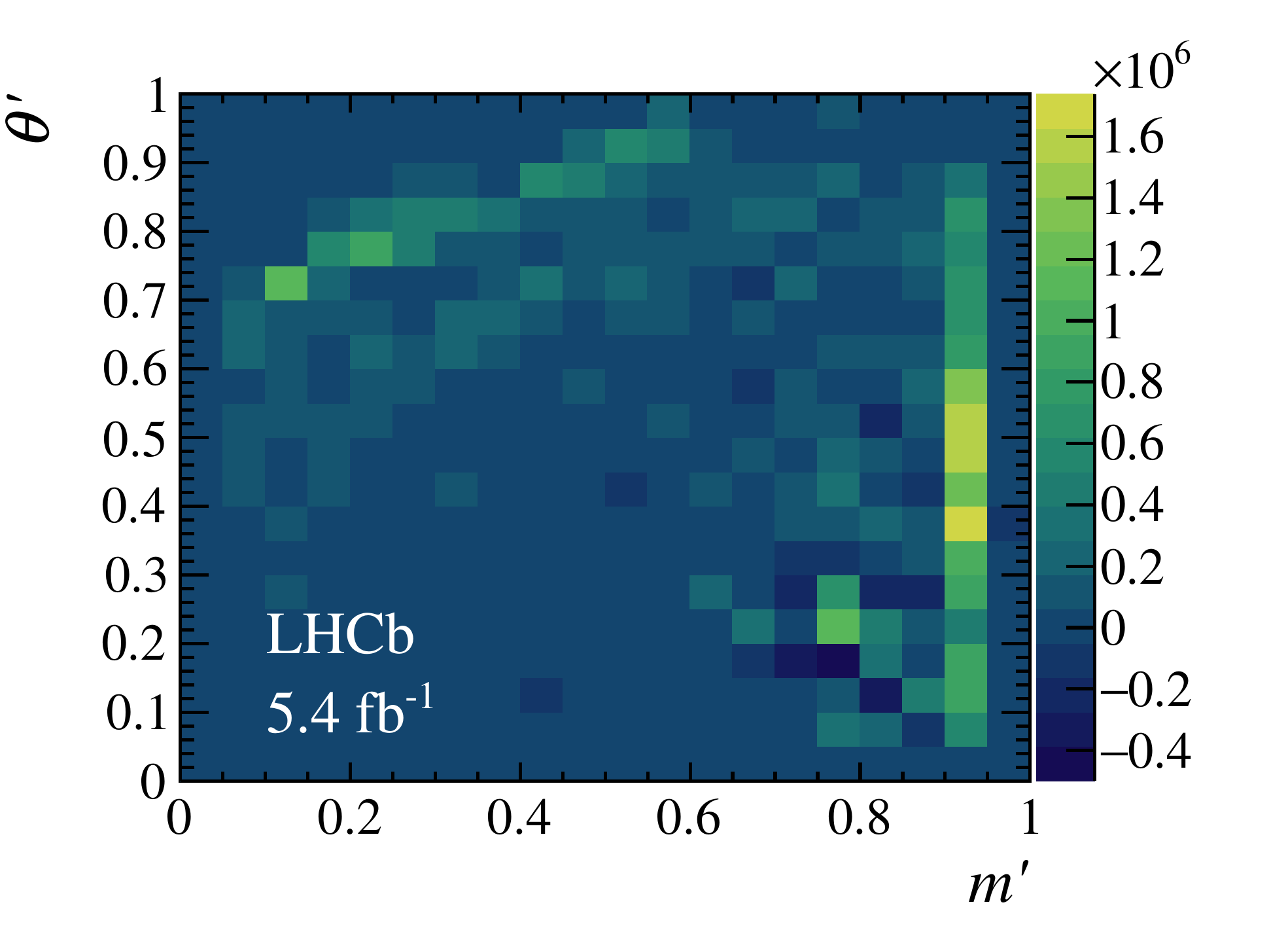}
  \includegraphics[width=0.49\textwidth]{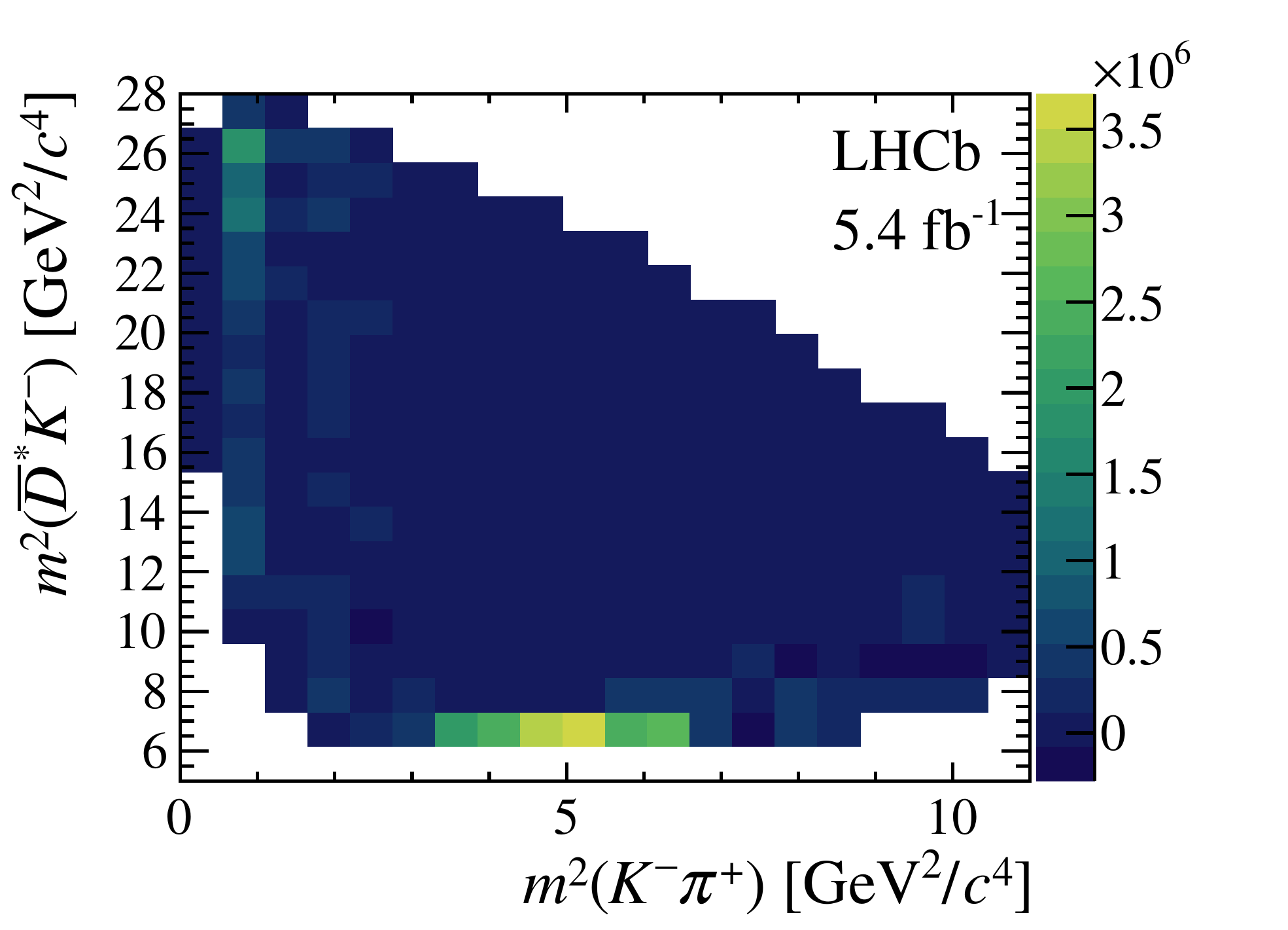}
  \includegraphics[width=0.49\textwidth]{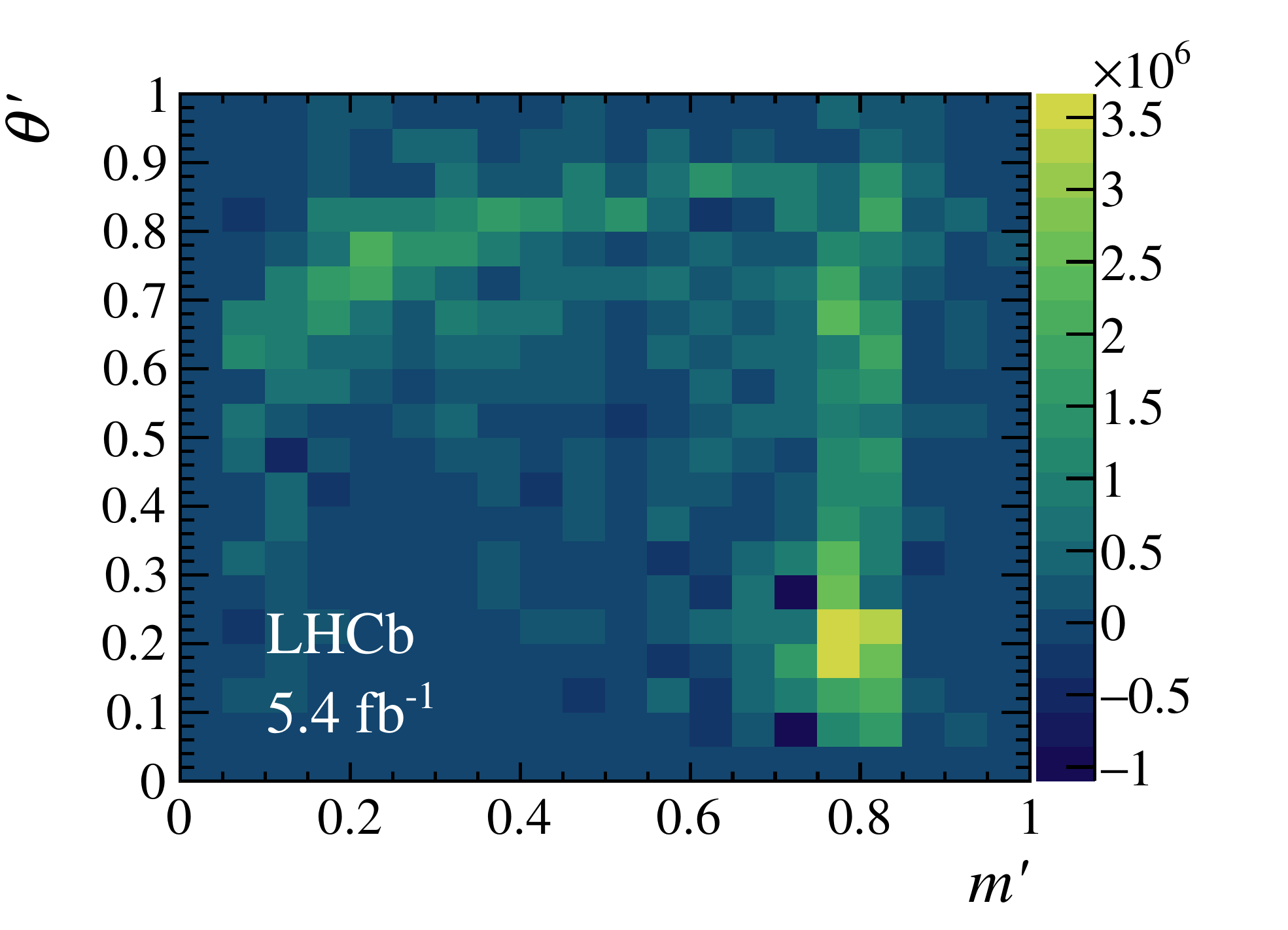}
  \includegraphics[width=0.49\textwidth]{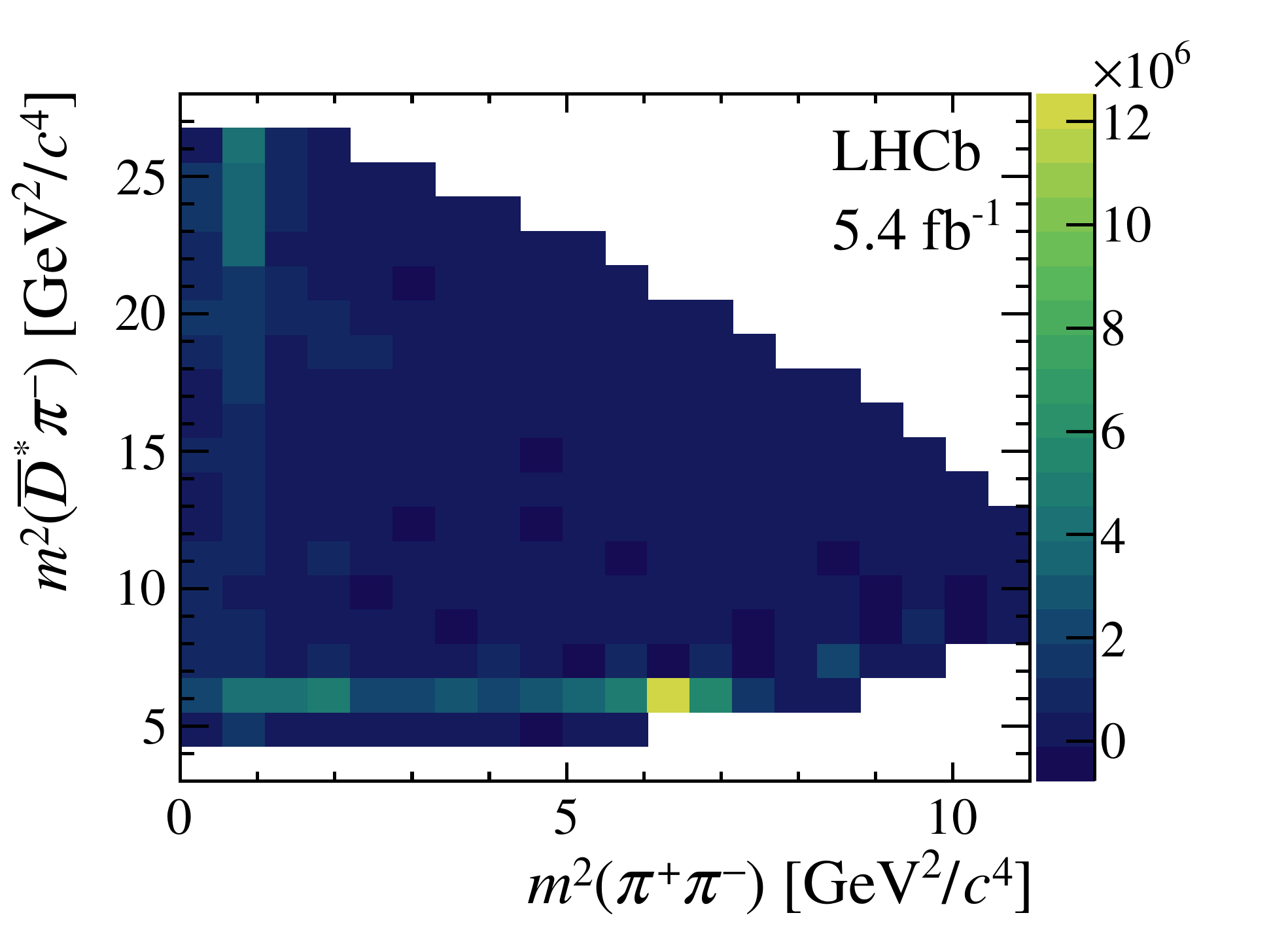}
  \caption{
    (Left)~SDP and (right)~conventional Dalitz plots of background-subtracted and efficiency-corrected data, for (top)~$\Bz\to\Dstarzb \Kp \pim$, (middle)~$\Bs\to\Dstarzb \Km \pip$ and (bottom)~$\Bz\to\Dstarzb \pip \pim$ decays.
  }
  \label{fig:DataDalitz}
\end{figure}

By plotting the efficiency-corrected signal weights as a function of position in phase space, rather than simply summing them as in Eq.~\eqref{eq:effWeight}, (square) Dalitz plots of the signal decays can be obtained.
These are shown in \cref{fig:DataDalitz} for (top)~$\Bz\to\Dstarzb \Kp \pim$, (middle)~$\Bs\to\Dstarzb \Km \pip$ and (bottom)~$\Bz\to\Dstarzb \pip \pim$ decays, for the $\Dstarzb\to\Dzb\gamma$ and $\Dstarzb\to\Dzb\piz$ samples combined.
Clear $K^\pm\pi^\mp$ structures, corresponding to the $\KorKbar{}^{*}(892)^0$ resonance, can be seen in both the $\Bz\to\Dstarzb\Kp\pim$ and $\Bs\to\Dstarzb\Km\pip$ channels.
Similarly, $\pip\pim$ structure corresponding to the $\rho(770)^0$ resonance can be seen in the normalisation channel.
Narrow $\Dstarzb h'^-$ structures can also be seen in all channels, corresponding to the $D_1(2420)^-$ resonance in $\Bz\to\Dstarzb\Kp\pim$ and $\Bz\to\Dstarzb \pip \pim$ decays, and corresponding to the $D_{s1}(2536)^-$ resonance in $\Bs\to\Dstarzb\Km\pip$ decays.
The structures observed are reasonably consistent with those anticipated when developing models for these channels to describe their shapes when appearing as misidentified background as discussed in Sec.~\ref{sec:massfits}. 
This provides confidence that the shapes used do not introduce significant bias into the results.

The signal weights can also be used to obtain distributions of variables used in the selection procedure, and thereby to verify that the simulation accurately represents the detector response and the signal kinematics.
Good agreement is found for all such variables.
The largest discrepancy is observed for one of the variables that quantifies the \BdorBs-candidate isolation from other activity in the $pp$ collision event.
This is considered as a source of systematic uncertainty on the results.

\section{Systematic uncertainties}
\label{sec:systematics}

A number of sources of possible bias on the ratios of branching fractions, evaluated using Eq.~\eqref{eq:effWeight}, are investigated and corresponding systematic uncertainties assigned.
These are  described below and summarised in the same order in \cref{tab:systematics}, where the statistical uncertainties are also given for comparison.
Separate measurements of the branching fraction ratios are obtained for the $\Dstarzb\to\Dzb\gamma$ and $\Dstarzb\to\Dzb\piz$ decays, and therefore separate sets of systematic uncertainties are presented.
These are indicated as being considered to be either completely uncorrelated ($\dagger$) or completely correlated ($*$) between the ratios obtained with the two \Dstarzb\ decays.
The correlated and uncorrelated sources are treated differently in the combination of results between the two channels, as discussed in Sec.~\ref{sec:results}.

\begin{table}[!tb]
  \caption{
    Systematic uncertainties (\%) on the branching fraction ratios, relative to the central values, with the soft neutral particle emitted in the \Dstarzb\ decay indicated as a subscript for brevity of notation.
    The sources of uncertainty are presented in the same order as described in the text, with statistical uncertainties included to facilitate comparison.
    Systematic uncertainties are indicated as being considered to be either completely uncorrelated ($\dagger$) or completely correlated ($*$) between the ratios obtained with the $\Dstarzb\to\Dzb\gamma$ and $\Dstarzb\to\Dzb\piz$ decays.
    The total correlated and uncorrelated systematic uncertainties are obtained by summing the relevant sources in quadrature.
    The uncertainty due to the fragmentation fractions is quoted separately.
  }
  \centering
  \resizebox{\textwidth}{!}{
    \def\arraystretch{1.25}
    \begin{tabular}{c|cccccc}
      \hline \\ [-3.3ex]
      & $\frac{\mathcal{B}(\Bz\to\Dbar{}^{*0}_{\gamma}\Kp\pim)}{\mathcal{B}(\Bz\to\Dbar{}^{*0}_{\gamma}\pip\pim)}$ & $\frac{\mathcal{B}(\Bz\to\Dbar{}^{*0}_{\piz}\Kp\pim)}{\mathcal{B}(\Bz\to\Dbar{}^{*0}_{\piz}\pip\pim)}$ & $\frac{\mathcal{B}(\Bs\to\Dbar{}^{*0}_{\gamma}\Km\pip)}{\mathcal{B}(\Bz\to\Dbar{}^{*0}_{\gamma}\pip\pim)}$ & $\frac{\mathcal{B}(\Bs\to\Dbar{}^{*0}_{\piz}\Km\pip)}{\mathcal{B}(\Bz\to\Dbar{}^{*0}_{\piz}\pip\pim)}$ & $\frac{\mathcal{B}(\Bz\to\Dbar{}^{*0}_{\gamma}\Kp\pim)}{\mathcal{B}(\Bs\to\Dbar{}^{*0}_{\gamma}\Km\pip)}$ &$\frac{\mathcal{B}(\Bz\to\Dbar{}^{*0}_{\piz}\Kp\pim)}{\mathcal{B}(\Bs\to\Dbar{}^{*0}_{\piz}\Km\pip)}$ \\ [1.0ex]
      \hline
      $\sigma_{\rm stat}$ & $6.1$  & $8.1$ & $2.7$  & $4.5$ & $5.9$  & $7.4$ \\
      \hline
      $\sigma^{*}_{\rm fit~bias}$ & $0.6$  & $0.8$ & $0.3$  & $0.1$ & $0.9$  & $1.0$ \\
      $\sigma^{*}_{\rm model}$ & $5.6$  & $3.8$ & $4.9$  & $2.1$ & $0.6$  & $6.0$ \\
      $\sigma^{*}_{\Lb}$ & $0.3$  & $1.5$ & $2.5$  & $4.4$ & $2.3$  & $2.7$ \\
      $\sigma^{*}_{\rm mult~cand}$ & $2.4$  & $2.5$ & $5.0$  & $0.2$ & $7.8$  & $2.7$ \\
      $\sigma^{\dagger}_{\rm MC~stats}$ & $1.2$  & $5.7$ & $1.2$  & $5.5$ & $1.4$  & $7.1$ \\
      $\sigma^{*}_{\rm PID}$ & $0.1$  & $0.0$ & $0.3$  & $0.2$ & $0.2$  & $0.2$ \\
      $\sigma^{*}_{\rm data/MC}$ & $0.1$  & $3.4$ & $0.6$  & $2.4$ & $0.5$  & $1.0$ \\
      $\sigma^{*}_{\rm trigger}$ & $1.1$  & $0.4$ & $0.4$  & $1.2$ & $0.7$  & $0.8$ \\
      $\sigma^{\dagger}_{\rm SDP~bins}$ & $1.4$  & $7.6$ & $1.9$  & $6.9$ & $2.1$  & $6.7$ \\
      $\sigma^{*}_{\rm 4D~PHSP}$ & $2.9$ & $4.0$ & $0.5$ & $0.8$ & $2.4$ & $3.1$  \\
      $\sigma^{*}_{\tau_\Bs}$ & --- & --- & $0.6$ & $0.5$ & $0.6$ & $0.5$ \\
      $\sigma^{*}_{\rm weights}$ & $0.6$ & $0.6$ & $1.0$ & $1.0$ & $1.6$ & $1.6$ \\
      \hline
      $\sigma_{\rm syst}^{\rm \dagger~total}$ & $1.8$  & $9.5$ & $2.3$  & $8.8$ & $2.5$  & $9.8$ \\
      $\sigma_{\rm syst}^{\rm *~total}$ & $6.9$  & $7.3$ & $7.7$  & $5.4$ & $8.6$  & $8.8$ \\
      \hline
      $\sigma^{*}_{f_s/f_d}$ & --- & --- & $3.1$  & $3.1$ & $3.1$  & $3.1$ \\ [0.3ex]
      \hline
    \end{tabular}
  }
  \label{tab:systematics}
\end{table}

\paragraph{Fit bias:}
Intrinsic bias in the simultaneous fit used to determine the signal yields is evaluated with an ensemble of pseudoexperiments.
Differences between the input values and the means of the fitted yields obtained in the ensemble, which are found to be small compared to the statistical uncertainties, are assigned as systematic uncertainties.
This procedure also reveals that the statistical uncertainty returned from the fit is underestimated by $10$--$30\%$, depending on the mode.
The statistical uncertainties are corrected accordingly (these corrections are included in the values presented in \cref{tab:systematics}).

\paragraph{\boldmath Modelling of components in the fit to \BdorBs-candidate mass:}
The largest uncertainty due to the choice of models used in the fit to \BdorBs-candidate mass arises from the reweighting of signal and misidentified background components.
This is investigated by varying the weights and evaluating the impact on the results.
The variation of weights accounts for both limited simulation size and uncertainties in the misidentification probabilities.

\paragraph{\boldmath Selection of components in the fit to \BdorBs-candidate mass:}
The baseline fit includes the components that are expected to contribute significantly.
However, contributions from $\Lbbar \to \Dstarzb \antiproton \Kp$ and $\Lbbar \to \Dstarzb \antiproton \pip$ decays, which are not included in the baseline fit, may be non-negligible.
To evaluate the possible impact of these background processes, the fit is repeated including these new components, with constraints relating their yields in the different final states.
The differences compared to the baseline results are assigned as the corresponding systematic uncertainty. 

\paragraph{Multiple and duplicate candidates:}
The effects of multiple and duplicate candidates may differ between data and simulation and could therefore bias the results.
This is investigated by imposing an additional selection requirement that reduces the rates of such candidates to zero, and evaluating the impact on the results.

\paragraph{Finite simulation statistics:}
The finite size of the simulation samples is a source of uncertainty in the efficiency estimation.
The calculation of each branching fraction ratio is repeated many times varying the efficiency in each SDP bin within the corresponding error bars.
The spread of results within this ensemble is assigned as the associated systematic uncertainty. 

\paragraph{Charged hadron identification:}
The sampling of variables related to charged hadron identification from data control samples requires that the detector response in the signal and control samples is the same.
This is expected to be the case, since dependence on kinematics and detector occupancy is accounted for in the procedure.
The impact of possible residual inconsistencies between the samples is evaluated by using an alternative sampling procedure, with the differences in the results obtained assigned as the associated systematic uncertainty. 

\paragraph{Disagreement between data and simulation in variables used in the selection:}
The weighting procedure described in Sec.~\ref{sec:efficiency} is used to check the agreement between data and simulation of variables used in the selection procedure.
The largest discrepancy is found in a variable that quantifies the \BdorBs-candidate isolation from other activity in the $pp$ collision event, which is used in one of the BDT classifiers.  
The BDT output is recalculated after reweighting this variable in simulation to match data, which affects the selection efficiencies. 
The impact on the results is assigned as the corresponding systematic uncertainty.  
  
\paragraph{Simulation of hardware trigger response:}
The response of the hardware trigger, due to energy deposits in the calorimeter system, is difficult to describe accurately in simulation.
The hardware trigger efficiency obtained in simulation is compared to that in data control samples, weighted to match the kinematics of signal tracks, and the impact on the results is assigned as the corresponding systematic uncertainty. 

\paragraph{SDP binning:}
The variation of the efficiency across the phase space is accounted for using binned SDP efficiency maps, and assumes that the variation of the efficiency within each bin is small enough to be neglected.
The possible bias related to this assumption is quantified by repeating the procedure with alternative SDP binnings.

\paragraph{Four-dimensional phase space:}
The phase space of a $\BdorBs \to \Dstarzb h^+h^{\prime -}$ decay has four dimensions.
The dependence of the efficiency on the SDP variables is accounted for in the baseline procedure.
The other two dimensions can be described by the cosine of the helicity angle of the \Dstarzb\ decay, $\cos \theta_{\Dstar}$, and the azimuthal angle, $\phi$, between the $\Dstarzb \to \Dzb\gamma/\piz$ decay plane and the plane defined by the $h^+$ and $h^{\prime -}$ momentum vectors.
The signal distributions in these variables are expected to be close to symmetric about zero and therefore if the efficiency dependence is approximately linear its impact on the results should be small.

The effect of the efficiency variation in $\cos \theta_{\Dstar}$ and $\phi$ is quantified by evaluating the average efficiency with and without efficiency corrections as functions of these variables.
The dependence on the SDP variables is omitted in this procedure, as the limited simulation sample size precludes a full four-dimensional treatment.
Effects related to $\phi$ are found to be negligible, as the efficiency has no significant dependence on this variable.
However, correlation between $\cos \theta_{\Dstar}$ with the momentum of the $\gamma$ or $\piz$ emitted in the $\Dstarzb$ decay results in a shift of the average efficiency, which is assigned as a systematic uncertainty.

\paragraph{\boldmath Data/simulation disagreement in \Bs\ lifetime:}
The $\Bs\to\Dstarzb\Km\pip$ decay is flavour-specific, and as such the appropriate combination of the lifetimes of the two physical \Bs\ eigenstates is the so-called flavour-specific lifetime $\tau_{\rm fs} = 1.527 \pm 0.011 \ps$~\cite{HFLAV18}.
The lifetime used when generating the simulation differs from this value by an amount that is approximately equal to the uncertainty on $\tau_{\rm fs}$.
Weights are applied to the simulation so that the \Bs-candidate decay-time distribution corresponds to $\tau_{\rm fs}$, and the impact on the result is assigned as the associated uncertainty.

\paragraph{Correlations that could bias the signal weighting procedure:}
The signal weighting procedure assumes the absence of correlation between the SDP variables and the \BdorBs-candidate mass.
This is an excellent approximation for signal, since the SDP coordinates are calculated with a \BdorBs-mass constraint applied, but correlations are expected for misidentified background and partially combinatorial background components.
The impact of these correlations is evaluated with pairs of pseudoexperiments in which these background sources are incorporated either with or without correlations between the \BdorBs-candidate mass and SDP variables.
The shift in results between the two cases is assigned as the associated systematic uncertainty.

\paragraph{Fragmentation fractions:}
Determination of the branching fraction ratios \mbox{${\cal B}(\Bs\rightarrow\Dstarzb\Km\pip)/{\cal B}(\Bz\rightarrow\Dstarzb\pip\pim)$} and \mbox{${\cal B}(\Bs\rightarrow\Dstarzb\Km\pip)/{\cal B}(\Bz\rightarrow\Dstarzb\Kp\pim)$} requires knowledge of the ratio of fragmentation fractions $f_s/f_d$.
The uncertainty in the knowledge of this quantity~\cite{LHCb-PAPER-2020-046} is therefore a source of systematic uncertainty.

\section{Results}
\label{sec:results}

The branching fraction ratios are calculated using Eq.~\eqref{eq:effWeight}, separately for cases with $\Dstarzb\to\Dzb\gamma$ and $\Dstarzb\to\Dzb\piz$ decays, resulting in the values presented in \cref{tab:results}.
The calculation of the statistical uncertainties accounts for the correlations given in \cref{tab:fitcorrelations}, assuming that these are also appropriate for the efficiency-corrected yields.
In all cases the same ratios evaluated with different \Dstarzb\ decays are found to be consistent.

\begin{table}[!tb]
  \centering
  \caption{
    Branching fraction ratios determined separately for $\Dstarzb\to\Dzb\gamma$ and $\Dstarzb\to\Dzb\piz$ decays. 
    The uncertainties are statistical, systematic and (where given) due to $f_s/f_d$.
  }
  \label{tab:results}
  \def\arraystretch{1.4}
  \begin{tabular}{ccc}
    \hline
    Ratio & $\Dstarzb\to\Dzb\gamma$ & $\Dstarzb\to\Dzb\piz$ \\
    \hline
    $\frac{\mathcal{B}(\Bz\rightarrow\Dstarzb\Kp\pim)}{\mathcal{B}(\Bz\rightarrow\Dstarzb\pip\pim)}$ & $0.083 \pm 0.005 \pm 0.008$ & $0.085 \pm 0.006\pm 0.012$ \\
    $\frac{\mathcal{B}(\Bs\rightarrow\Dstarzb\Km\pip)}{\mathcal{B}(\Bz\rightarrow\Dstarzb\pip\pim)}$ & $1.19 \pm 0.03 \pm 0.10\pm 0.04$ & $1.10 \pm 0.04 \pm 0.13\pm 0.03$ \\
    $\frac{\mathcal{B}(\Bz\rightarrow\Dstarzb\Kp\pim)}{\mathcal{B}(\Bs\rightarrow\Dstarzb\Km\pip)}$ & $0.070 \pm 0.004 \pm 0.008 \pm 0.002$ & $0.077 \pm 0.006 \pm 0.011\pm 0.002$ \\ [0.3ex]
    \hline
  \end{tabular}
\end{table}
    
A linear combination is used to average the two results for each branching fraction ratio.
The relative weight of each is determined using the standard minimum $\chisq$ approach, where only the statistical and uncorrelated systematic uncertainties are included.
These weights are then used to propagate the full set of uncertainties, including those that are correlated between the $\Dstarzb\to\Dzb\gamma$ and $\Dstarzb\to\Dzb\piz$ decays.
The results are 
\begin{eqnarray*}
  \frac{\mathcal{B}(\Bz\rightarrow\Dstarzb\Kp\pim)}{\mathcal{B}(\Bz\rightarrow\Dstarzb\pip\pim)} & = & 0.0836 \pm 0.0043 \pm 0.0061 \, , \\
  \frac{\mathcal{B}(\Bs\rightarrow\Dstarzb\Km\pip)}{\mathcal{B}(\Bz\rightarrow\Dstarzb\pip\pim)} & = & 1.178 \pm 0.029 \pm 0.091 \pm 0.037 \, , \\
  \frac{\mathcal{B}(\Bz\rightarrow\Dstarzb\Kp\pim)}{\mathcal{B}(\Bs\rightarrow\Dstarzb\Km\pip)} & = & 0.0712 \pm 0.0035 \pm 0.0065 \pm 0.0022 \, , 
\end{eqnarray*}
where the uncertainties are statistical, systematic and (where given) due to $f_s/f_d$.
The significance of the signal for each of the $\Bd\to\Dstarzb\Kp\pim$ and $\Bs\to\Dstarzb\Km\pip$ decays, as seen both in \cref{tab:FitResults} and in the results above, is far in excess of the range in which quantification of this value is useful, \ie\ it is $\gg 5\sigma$.

Using the previous measurement, $\mathcal{B}(\Bd\to\Dstarzb\pip\pim)=(6.2\pm1.2\pm1.8)\times10^{-4}$~\cite{Satpathy:2002js}, the absolute branching fractions of $\BdorBs\to\Dstarzb\Kpm\pimp$ decays are
\begin{eqnarray*}
    \mathcal{B}(\Bz\rightarrow\Dstarzb\Kp\pim) & = & (5.18 \pm 0.27 \pm 0.38 \pm 1.84 ) \times 10^{-5}\, , \\
    \mathcal{B}(\Bs\rightarrow\Dstarzb\Km\pip) & = & (7.30 \pm 0.18 \pm 0.56 \pm 2.59 \pm 0.23 ) \times 10^{-4}\, ,
\end{eqnarray*}
where the uncertainties are statistical, systematic, due to $\mathcal{B}(\Bd\to\Dstarzb\pip\pim)$ and (where given) to $f_s/f_d$.

In summary, the $\Bs\to\Dstarzb\Km\pip$ and $\Bd\to\Dstarzb\Kp\pim$ decays have been investigated, using a data sample corresponding to $5.4 \invfb$ of $pp$ collisions at centre-of-mass energy of $13\tev$.
Both decays are observed for the first time and their branching fractions are measured relative to that of the $\Bz\rightarrow\Dstarzb\pip\pim$ decay.
The knowledge obtained in this work of the branching fractions of these decays, and their distributions over phase space as shown in \cref{fig:DataDalitz}, will be important to control systematic uncertainties in future determinations of the CKM angle $\gamma$ in $B \to DK$ and $B \to DK\pi$ decays.
The $\Bz\rightarrow\Dstarzb\Kp\pim$ decay may itself also be used in future to obtain constraints on $\gamma$.
Moreover, future amplitude analyses of the Dalitz plots of these decays may provide insights into charm and charm-strange spectroscopy.

\section*{Acknowledgements}
%
%
\noindent We express our gratitude to our colleagues in the CERN
accelerator departments for the excellent performance of the LHC. We
thank the technical and administrative staff at the LHCb
institutes.
We acknowledge support from CERN and from the national agencies:
CAPES, CNPq, FAPERJ and FINEP (Brazil); 
MOST and NSFC (China); 
CNRS/IN2P3 (France); 
BMBF, DFG and MPG (Germany); 
INFN (Italy); 
NWO (Netherlands); 
MNiSW and NCN (Poland); 
MEN/IFA (Romania); 
MSHE (Russia); 
MICINN (Spain); 
SNSF and SER (Switzerland); 
NASU (Ukraine); 
STFC (United Kingdom); 
DOE NP and NSF (USA).
We acknowledge the computing resources that are provided by CERN, IN2P3
(France), KIT and DESY (Germany), INFN (Italy), SURF (Netherlands),
PIC (Spain), GridPP (United Kingdom), RRCKI and Yandex
LLC (Russia), CSCS (Switzerland), IFIN-HH (Romania), CBPF (Brazil),
PL-GRID (Poland) and NERSC (USA).
We are indebted to the communities behind the multiple open-source
software packages on which we depend.
Individual groups or members have received support from
ARC and ARDC (Australia);
AvH Foundation (Germany);
EPLANET, Marie Sk\l{}odowska-Curie Actions and ERC (European Union);
A*MIDEX, ANR, IPhU and Labex P2IO, and R\'{e}gion Auvergne-Rh\^{o}ne-Alpes (France);
Key Research Program of Frontier Sciences of CAS, CAS PIFI, CAS CCEPP, 
Fundamental Research Funds for the Central Universities, 
and Sci. \& Tech. Program of Guangzhou (China);
RFBR, RSF and Yandex LLC (Russia);
GVA, XuntaGal and GENCAT (Spain);
the Leverhulme Trust, the Royal Society
 and UKRI (United Kingdom).



\addcontentsline{toc}{section}{References}
\bibliographystyle{LHCb}
\bibliography{main,standard,LHCb-PAPER,LHCb-CONF,LHCb-DP,LHCb-TDR}
  
\newpage
\centerline
{\large\bf LHCb collaboration}
\begin
{flushleft}
\small
R.~Aaij$^{32}$,
A.S.W.~Abdelmotteleb$^{56}$,
C.~Abell{\'a}n~Beteta$^{50}$,
F.~Abudin{\'e}n$^{56}$,
T.~Ackernley$^{60}$,
B.~Adeva$^{46}$,
M.~Adinolfi$^{54}$,
H.~Afsharnia$^{9}$,
C.~Agapopoulou$^{13}$,
C.A.~Aidala$^{87}$
% ...

{\footnotesize \it

% ...
$^{9}$Universit{\'e} Clermont Auvergne, CNRS/IN2P3, LPC, Clermont-Ferrand, France\\
% ...
$^{13}$LPNHE, Sorbonne Universit{\'e}, Paris Diderot Sorbonne Paris Cit{\'e}, CNRS/IN2P3, Paris, France\\
% ...
$^{32}$Nikhef National Institute for Subatomic Physics, Amsterdam, Netherlands\\
% ...
$^{46}$Instituto Galego de F{\'\i}sica de Altas Enerx{\'\i}as (IGFAE), Universidade de Santiago de Compostela, Santiago de Compostela, Spain\\
% ...
$^{50}$Physik-Institut, Universit{\"a}t Z{\"u}rich, Z{\"u}rich, Switzerland\\
% ...
$^{54}$H.H. Wills Physics Laboratory, University of Bristol, Bristol, United Kingdom\\
% ...
$^{56}$Department of Physics, University of Warwick, Coventry, United Kingdom\\
% ...
$^{60}$Oliver Lodge Laboratory, University of Liverpool, Liverpool, United Kingdom\\
% ...
$^{68}$Syracuse University, Syracuse, NY, United States\\
% ...
$^{87}$University of Michigan, Ann Arbor, United States, associated to $^{68}$\\
\bigskip
% ...
}
\end{flushleft}

\end{document}